\documentstyle[sprocl,epsfig]{article}

\def\be{\begin{equation}}
\def\ee{\end{equation}}
\def\bea{\begin{eqnarray}}
\def\eea{\end{eqnarray}}

\begin{document}

\thispagestyle{empty}

\begin{flushright}
CERN-TH/2001-041
\end{flushright}

\vspace*{2cm}

\title{KAON AND CHARM PHYSICS: THEORY}

\author{G. BUCHALLA}

\address{Theory Division, CERN, CH-1211 Geneva 23, Switzerland\\
E-mail: Gerhard.Buchalla@cern.ch} 

\maketitle\abstracts{ 
We introduce and discuss basic topics in the theory of kaons
and charmed particles. In the first part, theoretical methods in
weak decays such as operator product expansion, renormalization
group and the construction of effective Hamiltonians are
presented, along with an elementary account of chiral perturbation
theory. The second part describes the phenomenology of the
neutral kaon system, CP violation, $\varepsilon$ and 
$\varepsilon'/\varepsilon$, rare kaon decays ($K\to\pi\nu\bar\nu$,
$K_L\to\pi^0 e^+e^-$, $K_L\to\mu^+\mu^-$), and some examples
of flavour physics in the charm sector.}

\vspace*{4cm}

\centerline{Lectures presented at TASI 2000, {\it Flavor Physics for
the Millennium},} 
\centerline{June 4--30, 2000, University of Colorado, Boulder, CO.}

\vfill

\begin{flushleft}
CERN-TH/2001-041
\end{flushleft}

\newpage
\pagenumbering{arabic}


\title{KAON AND CHARM PHYSICS: THEORY}

\author{G. BUCHALLA}

\address{Theory Division, CERN, CH-1211 Geneva 23, Switzerland\\
E-mail: Gerhard.Buchalla@cern.ch} 

\maketitle\abstracts{ 
We introduce and discuss basic topics in the theory of kaons
and charmed particles. In the first part, theoretical methods in
weak decays such as operator product expansion, renormalization
group and the construction of effective Hamiltonians are
presented, along with an elementary account of chiral perturbation
theory. The second part describes the phenomenology of the
neutral kaon system, CP violation, $\varepsilon$ and 
$\varepsilon'/\varepsilon$, rare kaon decays ($K\to\pi\nu\bar\nu$,
$K_L\to\pi^0 e^+e^-$, $K_L\to\mu^+\mu^-$), and some examples
of flavour physics in the charm sector.}

\section{Preface}

These lectures provide an introduction to the theory of
weak decays of kaons and mesons with charm.
Our main focus will be on kaon physics, which has led to
many deep and far-reaching insights into the structure of matter,
is a very active field of current research and still continues
to hold exciting opportunities for future discoveries.
Another, and in several ways complementary source of information
about flavour physics is the charm sector. Standard model effects
for rare processes are in this case typically suppressed to
almost negligible levels and positive signals, if observed,
could therefore yield spectacular evidence of new physics.
Towards the end of the lectures we will describe a few selected
examples in charm physics and contrast their characteristic features
with those of the kaon sector. For both subjects we will concentrate
on the flavour physics of the standard model. We discuss the
phenomenology as well as the theoretical tools necessary to
achieve a detailed and comprehensive test of the standard model
picture that should eventually lead us to uncover signals of
the physics beyond.
The experimental aspects of these fields are explained in the
lectures by Barker (kaon physics) and Cumalat
(charm physics) at this School.
Before we start our tour of flavour physics with kaons and charm,
we give a brief outline of the contents of these lectures.

We begin, in the following section 2, with recalling some of the
historical highlights of kaon physics and with an overview of
the main topics of current interest in this field.

In section 3 we introduce theoretical methods that are fundamental
for the computation of weak decay processes and for relating the
basic parameters of the underlying theory to actual observables.
These important tools are the operator product expansion, the
renormalization group, the effective low-energy weak Hamiltonians,
where we discuss the general $\Delta S=1$ Hamiltonian as an
explicit example, and, finally, chiral perturbation theory.
Our main emphasis will be on an elementary introduction of the
relevant ideas and concepts, rather than on more specialized
technical aspects.

With this background in mind we will then address, in section 4,
the phenomenology of the neutral-kaon system and CP violation.
We discuss a common classification of CP violation, the kaon
CP parameters $\varepsilon$ and $\varepsilon'/\varepsilon$, and
the standard analysis of the CKM unitarity triangle.

Section 5 is devoted to the physics of rare kaon decays, in particular 
the ``golden'' channels $K^+\to\pi^+\nu\bar\nu$ and 
$K_L\to\pi^0\nu\bar\nu$, and the processes $K_L\to\pi^0 e^+e^-$
and $K_L\to\mu^+\mu^-$.

In section 6 we discuss the prominent features of flavour
physics with charm. We present some opportunities with rare decays
of $D$ mesons and describe the phenomenology of $D^0$--$\bar D^0$
mixing, which is of current interest in view of new
experimental measurements by the CLEO and FOCUS collaborations.

Finally, section 7 summarizes the main points and presents an
outlook on future opportunities.

We conclude these preliminary remarks by mentioning several
review articles, which the interested reader may consult for
further details on the topics presented here, for a discussion
of related additional processes and for a complete collection
of references to the original literature.
Very useful accounts of rare and radiative kaon decays and
of kaon CP violation can be found in
\cite{LV,RW,WW,BR,BK,DI,AP}. The first five articles also discuss the
relevant experiments.
Nice reviews on flavour physics with charm are \cite{SP,GB,EG1,EG2}.
Further details on theoretical methods in weak decays are provided
in \cite{BBL,AJB1}.

\section{Kaons: Introduction and Overview}

\subsection{Historical Highlights}

The history of kaon physics is remarkably rich in
groundbreaking discoveries. It will be interesting to briefly
recall some of the most exciting examples here.
We do not attempt to give an historically accurate account of the
development of kaon physics. This is a fascinating subject in
itself. For a more complete historical picture the reader may
consult the excellent book by Cahn and Goldhaber \cite{CG}.
Here we will content ourselves with a brief sketch of several
highlights related to kaon physics. They serve to illustrate how
the observation of unexpected -- and sometimes tiny -- effects in this
field is linked to basic concepts in our theoretical understanding of
the fundamental interactions. 

\subsubsection{Strangeness}
Already the discovery of kaons alone, half a century ago, has had
an impressive impact on the development of high-energy physics.
One of the characteristic features of the new particles was
{\it associated production}, that is they were always produced in
pairs by strong interactions, for instance as

\begin{displaymath}
\begin{array}{cccccccccc}
& \pi^+ & + & p & \to & K^+ & + & \bar K^0 & + & p\\
{\rm S}\quad  & 0 & & 0 & & 
+1 & & -1 & & 0  
\end{array}
\end{displaymath}

\noindent
(Alternatively a $K^+$ could be produced along with a 
$\Lambda(uds)$ baryon.)
This property, together with the long lifetime, suggested
the existence of a new quantum number, called strangeness,
carried by the kaons and conserved in strong interactions.
The discovery of strangeness opened the way for the $SU(3)$
classification of hadrons and the introduction of quarks 
($u$, $d$, $s$) as the fundamental representation by Gell-Mann.
The quark picture, in turn, formed the basis for the subsequent
development of QCD.
In modern notation the $K$ mesons come in the following
varieties
\begin{tabbing}
\qquad\qquad\qquad\qquad\qquad\qquad 
\= $K^+ (\bar su)$\qquad \= $K^0(\bar sd)$ \\
\qquad\qquad\qquad\qquad\qquad\qquad 
\> $K^- (s\bar u)$\qquad \> $\bar K^0(s\bar d)$
\end{tabbing}
where the flavour content is indicated in brackets.
The pairs ($K^+$, $K^0$) and ($\bar K^0$, $K^-$) are doublets
of isospin.

\subsubsection{Parity Violation}

The new mesons proved to be strange particles indeed. One
of the peculiarities is known as the $\theta$--$\tau$ puzzle.
Two particles decaying as $\theta\to 2\pi$ (P even final state)
and $\tau\to 3\pi$ (P odd), and hence apparently
of different parity, were observed to have the same mass and lifetime.
This situation prompted Lee and Yang to propose that parity
might not be conserved in weak interactions. This was later confirmed
in the famous $^{60}{\rm Co}$ experiment by C.S. Wu.
Today the $\theta^+$ and $\tau^+$ are known to be identical
to the $K^+$ meson and parity violation is firmly encoded in the
chiral $SU(2)_L$ gauge group of standard model weak interactions.

\subsubsection{CP Violation}

After the recognition of parity violation in weak processes
the combination of parity with charge conjugation, CP, still appeared 
to be a good symmetry.
The neutral kaons $K^0$ and $\bar K^0$ were known to mix through
second order weak interactions to form, if CP was conserved,
the CP eigenstates $K_{L,S}=(K^0\pm \bar K^0)/\sqrt{2}$ (here
$CP\ K^0\equiv -\bar K^0$). Clearly, CP symmetry then forbids the
decay of the CP-odd $K_L$ into the CP-even $\pi^+\pi^-$ final state.
Instead, Christenson, Cronin, Fitch and Turlay showed in 1964 that
the decay does in fact occur, establishing CP violation.
Compared to the CP-allowed decay of $K_S\to\pi^+\pi^-$ the amplitude
is measured to be
\begin{equation}
\left|\frac{A(K_L\to\pi^+\pi^-)}{A(K_S\to\pi^+\pi^-)}\right|
= 2.3\cdot 10^{-3}\nonumber
\end{equation}
CP violation is thus a very small effect, in contrast to P violation,
but the {\it qualitative\/} implications are nevertheless 
far-reaching. As we will discuss later, CP violation defines
an absolute, and not only conventional, difference between matter
and anti-matter. Also, as we now know, CP violation indirectly
an\-ti\-ci\-pa\-ted in a sense the need for three families of fermions 
within the standard model. Finally, CP violation is a necessary
prerequisite for the generation of a net baryon number in
our universe according to Sakharov's three conditions (the other
two being baryon number violation and a departure from
thermal equilibrium).

\subsubsection{FCNC Suppression}

\begin{figure}[t]
\epsfig{figure=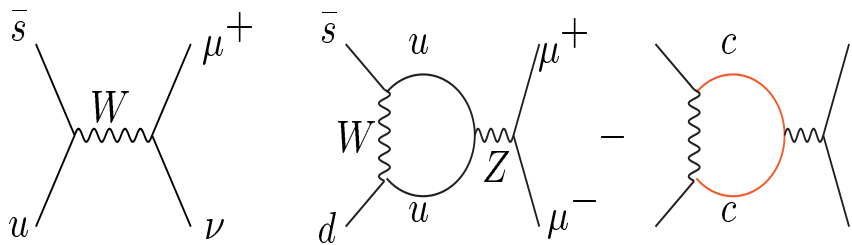,width=11.5cm,height=3.5cm}
\caption{GIM Mechanism.  \label{fig:gim}}
\end{figure}

Another striking property of weak interactions that manifested
itself in kaon decays is the  suppression of flavour-changing
neutral currents (FCNC).
While the standard, charged-current mediated process 
$K^+\to\mu^+\nu$ has a branching fraction of order unity
\begin{equation}\label{kpmn}
B(K^+\to\mu^+\nu)=0.64
\end{equation}
the similarly looking neutral-current decay $K_L\to\mu^+\mu^-$
is suppressed to a tiny level
\begin{equation}\label{klmm}
B(K_L\to\mu^+\mu^-)\approx 7\cdot 10^{-9}
\end{equation}
Naively, a ``three-quark standard model'' would allow a
$\bar sd Z$ coupling at tree level. This would lead to a 
$K_L\to\mu^+\mu^-$ amplitude of strength $G_F$, comparable
to $K^+\to\mu^+\nu$, in plain disagreement with (\ref{klmm}).
Even if the tree-level coupling of $\bar sd Z$ were forbidden,
the problem would reappear at one loop. This is illustrated
in the second diagram of Fig. \ref{fig:gim}.
The loop integral is divergent, where a natural cut-off could be
expected at the weak scale $\sim M_W$. The amplitude should then
be of the order $G^2_F M^2_W$, which would still be far too large.
Of course, the three-quark model is not renormalizable and therefore
not a consistent theory at short distances.
The introduction of the charm quark by Glashow, Iliopoulos and
Maiani (GIM) solves all of these problems, which plagued the early
theory of weak interactions. The complete two-generation standard
model is perfectly consistent and the tree-level $\bar sd Z$ coupling
is automatically eliminated by the orthogonality of the $2\times 2$
Cabibbo mixing matrix. The $\bar sd Z$ coupling can still be induced
at one-loop order, but the disturbing $G^2_F M^2_W$ term is now
canceled between the up-quark and the charm-quark contribution.
The remaining effect is, up to logarithms, merely of the order
$G^2_F m^2_c$, which is well compatible with (\ref{klmm}), unless
$m_c$ would be too large. To turn this observation into a more
quantitative constraint on the charm-quark mass $m_c$ is, however,
not easy in this case because $K_L\to\mu^+\mu^-$ is actually
dominated by long-distance contributions (we will discuss this 
further in section 5.3).
Another FCNC process, $K^0$--$\bar K^0$ mixing, proved to be
more useful in this respect.

\subsubsection{$K$--$\bar K$ mixing, GIM and Charm}

\begin{figure}[t]
\epsfig{figure=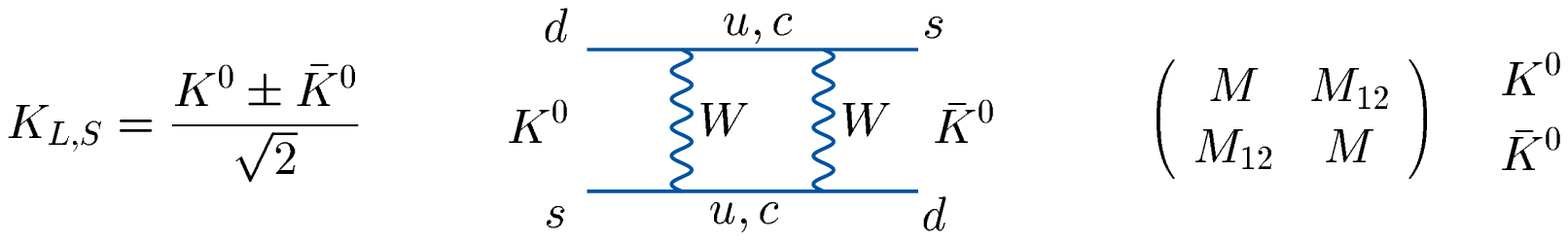,width=11.5cm,height=3.0cm}
\caption{$K^0$--$\bar K^0$ mixing.  \label{fig:kkmix}}
\end{figure}

The following example represents one of the great triumphs of early
standard model phenomenology. In the four-quark theory,
$K$--$\bar K$ mixing occurs through $\Delta S=2$ $W$-box diagrams
with internal up and charm quarks. This $\Delta S=2$ transition
induces a tiny off-diagonal element $M_{12}$ in the mass matrix
\begin{equation}\label{hm12}
H_M=\left(\begin{array}{cc}
              M & M_{12}\\
             M_{12} & M
\end{array}\right)
\end{equation}
of the $K$--$\bar K$ system. The corresponding eigenstates
are $K_{L,S}=(K^0\pm \bar K^0)/\sqrt{2}$ with eigenvalues
$M_{L,S}$. The difference between the eigenvalues
$\Delta M_K=M_L-M_S$ is related to $M_{12}$ and can be estimated 
from the box diagrams (see Fig. \ref{fig:kkmix}).
Anticipating a more detailed discussion of the calculation, we
simply quote the result:
\begin{equation}\label{dmk}
\frac{\Delta M_K}{M_K}\approx\frac{G^2_F f^2_K}{6\pi^2}
\left|V_{cs} V_{cd}\right|^2 m^2_c = 7\cdot 10^{-15}
\nonumber
\end{equation}
where the number on the right is the experimental value.
The theoretical expression is approximate since we have taken the
required hadronic matrix element in the so-called factorization
approximation, we have neglected the (fairly small) up-quark 
contribution and assumed that $m_c\ll M_W$.

Qualitatively, (\ref{dmk}) is easy to understand. There is a factor
of $G^2_F$ representing the second order weak interaction and a factor
of $f^2_K$ describing -- roughly speaking -- the bound state dynamics
binding the $s$ and $d$ quarks into kaons. The factor $6\pi^2$ in the
denominator is a typical numerical factor from the loop integration.
The whole amplitude is proportional to $m^2_c$, reflecting the remainder
of a GIM cancellation (assuming $m_c\gg\Lambda_{QCD}$, the up-quark
contribution is negligible). Finally there are the obvious CKM
parameters.

This analysis was performed in 1974 by Gaillard and Lee. The charm
quark had just been introduced for the theoretical reasons mentioned
above, but had not yet been discovered in experiment. Gaillard and Lee
realized that the quadratic dependence on the unknown
charm quark mass in (\ref{dmk}), resulting from the GIM cancellation
of the $\Delta S=2$ FCNC amplitude, could be turned into an
estimate of this mass. They concluded that $m_c$ should be about a few
GeV.
If you take the formula in (\ref{dmk}) and put in numbers, you will
find that $m_c\approx 1.5\,{\rm GeV}$ (!).
In view of the uncertainties of (\ref{dmk}) the accuracy of this result
cannot be taken too seriously, but it is in any case amusing that
indeed a quite realistic charm mass comes out already from this 
simplified formula.
What is more important, however, is the spirit of the argument,
which led Gaillard and Lee to a correct prediction of $m_c$ by taking
the short-distance structure of the theory seriously. This was
certainly one of the most beautiful successes of flavour physics with
kaons. 

All these examples illustrate that a careful analysis of low-energy
processes such as kaon decays, can lead to truly profound insights
into fundamental physics. Especially remarkable is the circumstance
that kaon physics, which ``operates'' at the $\sim 500\,{\rm MeV}$
scale, carries important information on the dynamics at much
higher energy scales. The quark-structure of hadronic matter, the
chiral nature of weak gauge interactions, the properties of charm,
and those of the top quark entering CP violating amplitudes, are
prominent examples that illustrate this point.
In fact, we see that several of the most crucial pillars of the
standard model rest on results derived from studies with kaons.
Of course, direct experiments at high energies, that have led for
instance to the production of on-shell $W$ and $Z$ bosons or quark
jets, are indispensable for exploring the strucutre of matter.
However, indirect, low-energy precision observables are equally
necessary as a complementary approach. They can yield information that is
hardly accessible in any other way, such as the elucidation of the GIM
structure of flavour physics or the violation of CP symmetry.
It is with this philosophy in mind that studies of rare kaon
processes, but also rare decays of $b$ hadrons or charmed particles,
continue to be pursued with great interest.  
The most promising future opportunities with kaons will be
the subject of later sections in these lectures.
We conclude this introductory chapter with a
brief general overview of physics with kaon decays.

\subsection{Overview of $K$ decays}

We may classify the decays of $K$ mesons into several broad
categories, some of which are more determined by nonperturbative
strong interaction dy\-na\-mics, while others have a high sensitivity
to short-distance physics both in the standard model and beyond.

\subsubsection{Tree-Level (Semi-) Leptonic Decays}

\begin{figure}[t]
\epsfig{figure=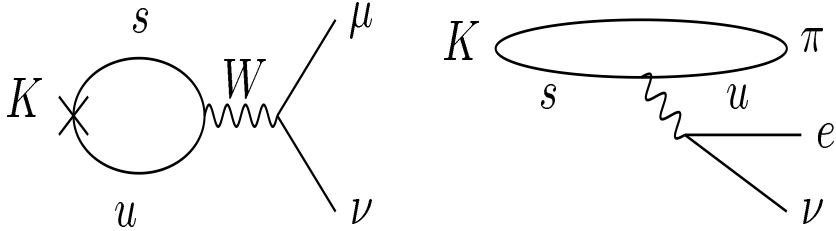,width=11.5cm,height=3.5cm}
\caption{(Semi)leptonic $K$ decays.  \label{fig:ksl}}
\end{figure}

These are the simplest decays of kaons. They typically have large branching
ratios and are well studied. Examples are the purely
leptonic decay $K^+\to\mu^+\nu$ and the semileptonic mode
$K^+\to\pi^0 e^+\nu$, which are illustrated in Fig. \ref{fig:ksl}.
$K^+\to\pi^0 e^+\nu$ is very important for determining the CKM
matrix element $V_{us}$ (the sine of the Cabibbo angle).
This is possible because the hadronic matrix element of the
vector current $\langle\pi^0|(\bar su)_V|K^+\rangle$ is absolutely
normalized in the limit of SU(3) flavour symmetry and protected
from first order corrections in the SU(3) breaking (Ademollo-Gatto theorem).
One finds
\begin{equation}\label{vusexp}
|V_{us}|=0.2196\pm 0.0023
\end{equation}
This is a basic input for the CKM matrix.
Furthermore, knowing $|V_{us}|$, $K^+\to\mu^+\nu$ may be used to
determine the kaon decay constant $f_K=160\,{\rm MeV}$.

\subsubsection{Nonleptonic Decays}

\begin{figure}[h]
\epsfig{figure=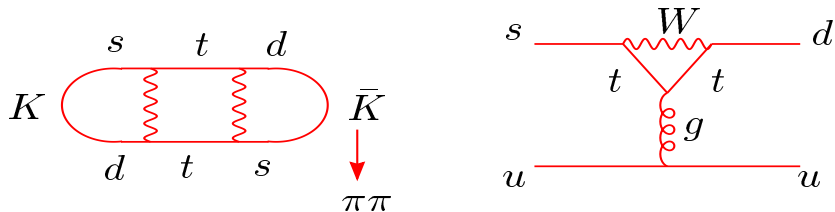,width=11.5cm,height=5cm}
\vspace*{-2cm}
\caption{SM origin of CP violation in $K\to\pi\pi$ decays. \label{fig:kppcp}}
\end{figure}

Nonleptonic decays, such as $K\to\pi\pi$, are strongly affected
by nonperturbative QCD dynamics. Nevertheless they provide an
important window on the violation of discrete symmetries, P and CP
for example. CP violation is currently of special interest. It enters
through $K$--$\bar K$ mixing via box graphs or through penguin
diagrams in the decay amplitudes, with the virtual top quarks playing
a decisive role. This is sketched in Fig. \ref{fig:kppcp}.

\subsubsection{Long-Distance Dominated Rare and Radiative Decays}

\begin{figure}[t]
\hspace*{3.5cm}\epsfig{figure=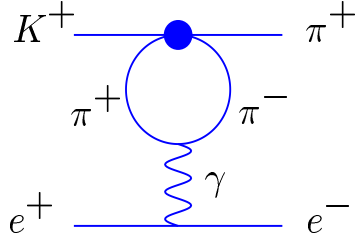,width=5cm,height=3.5cm}
\caption{Typical contribution to $K^+\to\pi^+ e^+e^-$. \label{fig:kppee}}
\end{figure}

Examples of this class of processes are 
$K^+\to\pi^+l^+l^-$, $K_L\to\pi^0\gamma\gamma$, $K_S\to\gamma\gamma$
or $K_L\to\mu^+\mu^-$.
A typical contribution to $K^+\to\pi^+ e^+e^-$ is illustrated
in Fig. \ref{fig:kppee}. 
These processes are determined by
nonperturbative low-energy strong interactions and can be analyzed
in the framework of chiral perturbation theory. The treatment of
nonperturbative dynamics within a first-principles approach as
provided by chiral perturbation theory is of great interest in its
own right. In addition, the control over long-distance contributions
afforded by chiral perturbation theory can be helpful to extract
information on the flavour physics from short distances.

\subsubsection{Short-Distance Dominated Rare Decays}

The prime examples in this category are the processes
$K^+\to\pi^+\nu\bar\nu$ and $K_L\to\pi^0\nu\bar\nu$. Here
short-distance dynamics completely dominates the decay and the
access to flavour physics is very clean and direct.
For this reason the $K\to\pi\nu\bar\nu$ modes are special
highlights among the future opportunities in kaon physics.
To a somewhat lesser extent also $K_L\to\pi^0 e^+e^-$ qualifies
for this class. In this case the fact that the process is predominantly
CP violating enhances the sensitivity to short-distance physics
in comparison to $K^+\to\pi^+ e^+e^-$.

\subsubsection{Decays Forbidden in the SM}

Any positive signal in processes that are forbidden in the standard
model would be a very dramatic indication of new physics. A good
example are kaon decays with lepton-flavour violation. Stringent
experimental limits exist for several modes of interest
\begin{eqnarray}
B(K_L\to\mu e) &<& 4.7\cdot 10^{-12} \label{lfvexp}\\
B(K^+\to\pi^+\mu^+ e^-) &<& 4.8\cdot 10^{-11} \\
B(K_L\to\pi^0\mu e) &<& 3.2\cdot 10^{-9} 
\end{eqnarray}
In principle these processes could be induced through loops in
the standard model with neutrino masses. However, the smallness of
the neutrino masses compared to the weak scale results in
unmeasurably small branching fractions of typically below $10^{-25}$.
Larger values can be obtained within the minimal supersymmetric
standard model (MSSM). However, there are strong constraints from direct
limits on $\mu\to e$ conversion processes
($\mu\to e\gamma$ decay, or $\mu\to e$ conversion in the field of a
nucleus). The disadvantage of $K_L\to\mu e$ is that flavour violation
is needed simultaneously both in the lepton sector and in the quark
sector. Interesting effects could however still occur in some regions
of parameter space. Systematically larger branching ratios are allowed
in scenarios with R-parity violation, where decays such as $K_L\to\mu e$
can proceed at tree level \cite{BCDEGL}. 

In very general terms, a scenario where the exchange of a heavy boson
$X$ mediates $\bar sd\to\bar\mu e$ transitions at tree level
receives strong constraints from the tight experimental bound 
(\ref{lfvexp}). Assuming couplings of electroweak strength,
the bound (\ref{lfvexp}) implies a lower limit of the $X$ mass
$M_X\stackrel{>}{_\sim} 100\,{\rm TeV}$. Such a sensitivity to
high energy scales is very impressive, however one has to remember
that the tree-level scenario assumed above is quite simple-minded and
in general very model dependent. Generically, one would expect some
additional suppression mechanism to be at work in $K_L\to\mu e$. In this
case the scale probed would be less, but the high precision of
(\ref{lfvexp}) still guarantees an excellent sensitivity to
subtle short-distance effects.

\section{Theoretical Methods in Weak Decays}

\begin{figure}[t]
\epsfig{figure=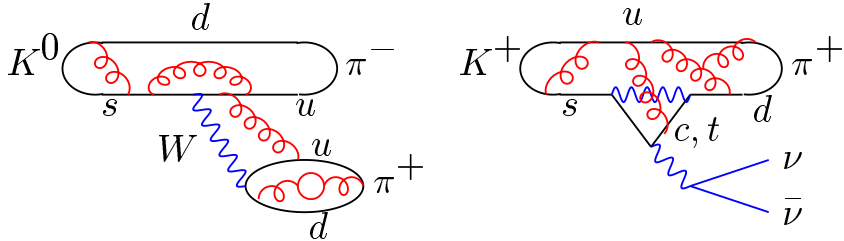,width=11.5cm,height=3.5cm}
\caption{QCD effects in weak decays. \label{fig:kppkpnn}}
\end{figure}

The task of computing weak decays of kaons represents a complicated
problem in quantum field theory.
Two typical cases, the first-order nonleptonic process
$K^0\to\pi^+\pi^-$, and the loop-induced, second-order weak
transition $K^+\to\pi^+\nu\bar\nu$ are illustrated in 
Fig. \ref{fig:kppkpnn}.
The dynamics of the decays is determined by a nontrivial interplay
of strong and electroweak forces, which is characterized by several
energy scales of very different magnitude, the $W$ mass, the various
quark masses and the QCD scale:
$m_t$, $M_W\gg$ $m_c\gg$ $\Lambda_{QCD} \gg$ $m_u$, $m_d$, $(m_s)$.
While it is usually sufficient to treat electroweak interactions
to lowest nonvanishing order in perturbation theory, it is necessary
to consider all orders in QCD. Asymptotic freedom still allows us to
compute the effect of strong interactions at short distances
perturbatively. However, since kaons are bound states of light quarks,
confined inside the hadron by long-distance dynamics, it is clear that 
also nonperturbative QCD interactions enter the decay process in an
essential way.

To deal with this situation, we need a method to disentangle long-
and short-distance contributions to the decay amplitude in a systematic
fashion. The required tool is provided by the operator product
expansion (OPE).

\subsection{Operator Product Expansion}

\begin{figure}[h]
\epsfig{figure=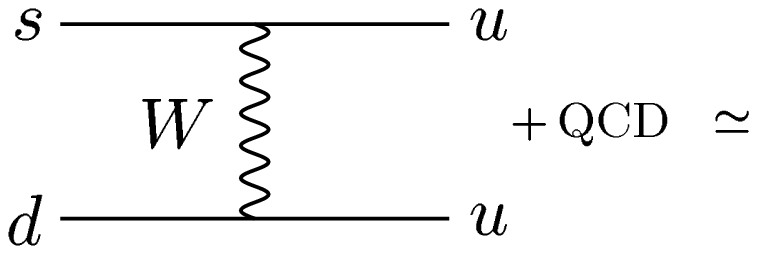,width=5cm,height=3.0cm}
\raisebox{1.3cm}{$C\left(\frac{M_W}{\mu}, \alpha_s\right)\cdot$}
\epsfig{figure=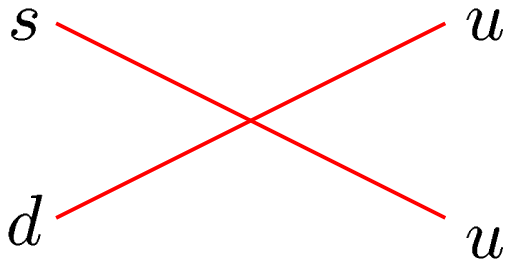,width=4cm,height=3.0cm}
\caption{OPE for weak decays. \label{fig:ope}}
\end{figure}

We will now discuss the basic concepts of the OPE for
kaon decay amplitudes. These concepts are of crucial importance
for the theory of weak decay processes, not only of kaons, but also
of mesons with charm and beauty and other hadrons as well.
Consider, for instance, the basic $W$-boson exchange process shown on 
the left-hand side of Fig. \ref{fig:ope}. 
This diagram mediates the
decay of a strange quark and triggers the nonleptonic decay of a kaon
such as $K^0\to\pi^+\pi^-$. The quark-level transition shown is
understood to be dressed with QCD interactions of all kinds, including
the binding of the quarks into the mesons. To simplify this
problem, we may look for a suitable expansion parameter, as we are
used to do in theoretical physics. Here, the key feature is provided
by the fact that the $W$ mass $M_W$ is very much heavier than the
other momentum scales $p$ in the problem 
($\Lambda_{QCD}$, $m_u$, $m_d$, $m_s$). We can therefore expand the
full amplitude $A$, schematically, as follows
\begin{equation}\label{acq}
A=C\left(\frac{M_W}{\mu},\alpha_s\right)\cdot\langle Q\rangle
+{\cal O}\left(\frac{p^2}{M^2_W}\right)
\end{equation}
which is sketched in Fig. \ref{fig:ope}. Up to negligible
power corrections of ${\cal O}(p^2/M^2_W)$, the full amplitude
on the left-hand side is written as the matrix element of a local
four-quark operator $Q$, multiplied by a Wilson coefficient $C$.
This expansion in $1/M_W$ is called a (short-distance) operator
product expansion because the nonlocal product of two bilinear
quark-current operators $(\bar su)$ and $(\bar ud)$ that interact
via $W$ exchange, is expanded into a series of local operators.
Physically, the expansion in Fig. \ref{fig:ope} means that the
exchange of the very heavy $W$ boson can be approximated by a
point-like four-quark interaction. With this picture the formal
terminology of the OPE can be expressed in a more intuitive language
by interpreting the local four-quark operator as a four-quark
interaction vertex and the Wilson coefficient as the corresponding
coupling constant. Together they define an effective Hamiltonian
${\cal H}_{eff}=C\cdot Q$, describing weak interactions of light
quarks at low energies.
Ignoring QCD the OPE reads explicitly (in momentum space)
\begin{eqnarray}\label{atree}
A &=& \frac{g^2_W}{8}V^*_{us}V_{ud}\frac{i}{k^2-M^2_W}
(\bar su)_{V-A}(\bar ud)_{V-A} \nonumber \\
&=& -i\frac{G_F}{\sqrt{2}}V^*_{us} V_{ud} C\cdot\langle Q\rangle
+{\cal O}\left(\frac{k^2}{M^2_W}\right)
\end{eqnarray}
with $C=1$, $Q=(\bar su)_{V-A}(\bar ud)_{V-A}$ and
\begin{equation}\label{htree}
{\cal H}_{eff}=\frac{G_F}{\sqrt{2}}V^*_{us} V_{ud}
(\bar su)_{V-A}(\bar ud)_{V-A}
\end{equation}

As we will demonstrate in more detail below after including QCD
effects, the most important property of the OPE in (\ref{acq})
is the {\it factorization \/} of long- and short-distance contributions:
All effects of QCD interactions above some factorization scale $\mu$
(short distances) are contained in the Wilson coefficient $C$.
All the low-energy contributions below $\mu$ (long distances) are
collected into the matrix elements of local operators $\langle Q\rangle$.
In this way the short-distance part of the amplitude can be 
systematically extracted and calculated in perturbation theory.
The problem to evaluate the matrix elements of local operators between
hadron states remains. This task requires in general nonperturbative
techniques, as for example lattice QCD, but it is considerably
simpler than the original problem of the full standard-model amplitude.
In some cases also symmetry considerations can help to determine
the nonperturbative input. For example, the only matrix element
relevant for $K^+\to\pi^+\nu\bar\nu$ is
\begin{equation}\label{kpiso}
\langle\pi^+|(\bar sd)_V|K^+\rangle =
\sqrt{2} \langle\pi^0|(\bar su)_V|K^+\rangle
\end{equation}
where the equality with the right-hand side uses isospin symmetry
and allows us to obtain the matrix element from measuring the
standard semileptonic mode $K^+\to\pi^0 l^+\nu$.

The short-distance OPE that we have described, the resulting
effective Hamiltonian, and the factorization property are fundamental
for the theory of $K$ decays. However, the concept of factorization
of long- and short-distance contributions reaches far beyond these
applications. In fact, the idea of factorization, in various forms
and generalizations, is the key to essentially all applications
of perturbative QCD, including the important areas of deep-inelastic
scattering and jet or lepton pair
production in hadron-hadron collisions. The reason is the same in
all cases: Perturbative QCD is a theory of quarks and gluons, but those
never appear in isolation and are always bound inside hadrons.
Nonperturbative dynamics is therefore always relevant to some extent
in hadronic reactions, even if these occur at very high energy or
with a large intrinsic mass scale (see also the lectures by Soper
in these proceedings). Thus, before perturbation theory can be
applied, nonperturbative input has to be isolated in a systematic
way, and this is achieved by establishing the property of 
factorization.
It turns out that the weak effective Hamiltonian for $K$ decays
provides a nice example to demonstrate the general idea of
factorization in simple and explicit terms.

\begin{figure}[t]
\epsfig{figure=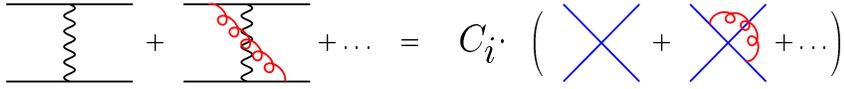,width=11.5cm,height=3.5cm}
\vspace*{-2cm}
\caption{Calculation of Wilson coefficients of the OPE. \label{fig:opeqcd}}
\end{figure}

We will next discuss the OPE for $K$ decays, now including the effects
of QCD, and illustrate the calculation of the Wilson coefficients.
A diagrammatic representation for the OPE is shown in 
Fig. \ref{fig:opeqcd}.
The key to calculating the coefficients $C_i$ is again the
property of factorization. Since factorization implies the separation
of all long-distance sensitive features of the amplitude into the
matrix elements of $\langle Q_i\rangle$, the short-distance quantities
$C_i$ are, in particular, independent of the external states.
This means that the $C_i$ are always the same, no matter whether we
consider the actual physical amplitude where the quarks are bound inside
mesons, or any other, unphysical amplitude with on-shell or even
off-shell external quark lines. Thus, even though we are ultimately
interested in $K\to\pi\pi$ amplitudes, for the perturbative
evaluation of $C_i$ we are free to choose any treatment of the
external quarks according to our calculational convenience.
A convenient choice that we will use below is to take all light
quarks massless and with the same off-shell momentum $p$ ($p^2\not= 0$).

The computation of the $C_i$ in perturbation theory then proceeds in
the following steps:

\begin{itemize}
\item 
Compute the amplitude $A$ in the full theory (with $W$ propagator)
for arbitrary external states.
\item
Compute the matrix elements $\langle Q_i\rangle$ with the same
treatment of external states.
\item
Extract the $C_i$ from $A=C_i\, \langle Q_i\rangle$.
\end{itemize}

\begin{figure}[t]
\hspace*{3.8cm}\epsfig{figure=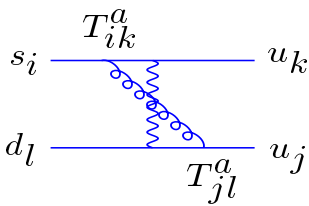,width=7cm,height=5cm}
\vspace*{-2cm}
\caption{QCD correction with colour assignment. \label{fig:q1col}}
\end{figure}

We remark that with the off-shell momenta $p$ for the quark lines
the amplitude is even gauge dependent and clearly unphysical.
However, this dependence is identical for $A$ and $\langle Q_i\rangle$
and drops out in the coefficients. The actual calculation is most
easily performed in Feynman gauge.
To ${\cal O}(\alpha_s)$ there are four relevant diagrams, the one
shown in Fig. \ref{fig:opeqcd} together with the remaining three
possibilities to connect the two quark lines with a gluon.
Gluon corrections to either of these quark currents need not be
considered, they are the same on both sides of the OPE and drop out
in the $C_i$.
The operators that appear on the right-hand side follow from the
actual calculations.
Without QCD corrections there is only one operator of dimension 6
\begin{equation}\label{q2def}
Q_2=(\bar s_i u_i)_{V-A}(\bar u_j d_j)_{V-A}
\end{equation}
where the colour indices have been made explicit.
(The operator is termed $Q_2$ for historical reasons.)
To ${\cal O}(\alpha_s)$ QCD generates another operator
\begin{equation}\label{q1def}
Q_1=(\bar s_i u_j)_{V-A}(\bar u_j d_i)_{V-A}
\end{equation}
which has the same Dirac and flavour structure, but a different
colour form.
Its origin is illustrated in Fig. \ref{fig:q1col},
where we recall the useful identity for $SU(N)$ Gell-Mann matrices
\begin{equation}\label{tata}
(\bar s_i T^a_{ik} u_k) (\bar u_j T^a_{jl} d_l)=
-\frac{1}{2N}(\bar s_i u_i) (\bar u_j d_j)
+\frac{1}{2}(\bar s_i u_j) (\bar u_j d_i)
\end{equation}
It is convenient to employ a different operator basis, defining
\begin{equation}\label{qpm}
Q_\pm=\frac{Q_2\pm Q_1}{2}
\end{equation}
The corresponding coefficients are then given by
\begin{equation}\label{cpm}
C_\pm=C_2\pm C_1
\end{equation}
If we denote by $S_\pm$ the spinor expressions that correspond
to the operators $Q_\pm$ (in other words: the tree-level matrix
elements of $Q_\pm$), the full amplitude can be written as
\begin{equation}\label{aft}
A=\left(1+\gamma_+ \alpha_s \ln\frac{M^2_W}{-p^2}\right)S_+
+ \left(1+\gamma_- \alpha_s \ln\frac{M^2_W}{-p^2}\right)S_-
\end{equation}
Here we have focused on the logarithmic terms and dropped
a constant contribution (of order $\alpha_s$, but nonlogarithmic).
Further, $p^2$ is the virtuality of the quarks and $\gamma_\pm$
are numbers that we will specify later on.
We next compute the matrix elements of the operators in the effective
theory, using the same approximations, and find
\begin{equation}\label{qme}
\langle Q_\pm\rangle=
\left(1+\gamma_\pm \alpha_s\left(\frac{1}{\varepsilon}+
\ln\frac{\mu^2}{-p^2}\right)\right) S_\pm
\end{equation}
The divergence that appears in this case has been regulated in
dimensional regularization ($D=4-2\varepsilon$ dimensions).
Requiring
\begin{equation}\label{acqpm}
A=C_+ \langle Q_+\rangle + C_- \langle Q_-\rangle
\end{equation}
we obtain
\begin{equation}\label{cll}
C_\pm = 1+\gamma_\pm \alpha_s \ln\frac{M^2_W}{\mu^2}
\end{equation}
where the divergence has been subtracted in the minimal subtraction
scheme.
The effective Hamiltonian we have been looking for then reads
\begin{equation}\label{hcqpm}
{\cal H}_{eff}=\frac{G_F}{\sqrt{2}}V^*_{us}V_{ud}
\left( C_+(\mu) Q_+ + C_-(\mu) Q_-\right)
\end{equation}
with the coefficients $C_\pm$ determined in (\ref{cll}) to
${\cal O}(\alpha_s\,{\rm log})$ in perturbation theory.
The following points are worth noting:

\begin{itemize}
\item
The $1/\varepsilon$ (ultraviolet) divergence in the effective theory
(\ref{qme}) reflects the $M_W\to\infty$ limit. This can be seen from
the amplitude in the full theory (\ref{aft}),
which is finite, but develops a logarithmic singularity in this limit.
Consequently, the renormalization in the effective theory is directly
linked to the $\ln M_W$ dependence of the decay amplitude.
\item
We observe that although $A$ and $\langle Q_\pm\rangle$ both
depend on the long-distance properties of the external states
(through $p^2$), this 
de\-pen\-dence has drop\-ped out in $C_\pm$.
Here we see explicitly how factorization is realized.
Technically, to ${\cal O}(\alpha_s\,{\rm log})$, factorization is
equivalent to splitting the logarithm of the full amplitude 
according to
\begin{equation}
\ln\frac{M^2_W}{-p^2}=\ln\frac{M^2_W}{\mu^2}+\ln\frac{\mu^2}{-p^2}
\end{equation}
Ultimately the logarithms stem from loop momentum integrations
and the range of large momenta, between
$M_W$ and the factorization scale $\mu$, is indeed separated into the
Wilson coefficients.
\item
To obtain a decay amplitude from ${\cal H}_{eff}$ in
(\ref{hcqpm}), the matrix elements
$\langle f|Q_\pm|K\rangle(\mu)$ have to be taken, normalized at a
scale $\mu$. An appropriate value for $\mu$ is close to the hadronic
scale in order not to introduce an unnaturally large scale into
the calculation of $\langle Q\rangle$. At the same time $\mu$ must
also not be too small in order not to render the perturbative
calculation of $C(\mu)$ invalid. A typical choice for $K$ decays
is $\mu\approx 1\,{\rm GeV}\ll M_W$.
\item
The factorization scale $\mu$ is unphysical. It cancels between
Wilson coefficient and hadronic matrix element, to a given order
in $\alpha_s$, to yield a scale independent decay amplitude.
The mechanism of this cancellation to ${\cal O}(\alpha_s)$ is
clear from the above example (\ref{aft}) -- (\ref{cll}).
\item
In the construction of ${\cal H}_{eff}$ the $W$-boson is said to
be ``integrated out'', that is, removed from the effective theory as
an explicit degree of freedom. Its effect is still implicitly
contained in the Wilson coefficients. The extraction of these
coeffcients is often called a ``matching calculation'',
matching the full to the effective theory by ``adjusting'' the
couplings $C_\pm$.
\item
If we go beyond the leading logarithmic approximation
${\cal O}(\alpha_s\log)$ and include the finite corrections
of ${\cal O}(\alpha_s)$ in (\ref{aft}), (\ref{qme}), an ambiguity
arises when renormalizing the divergence in (\ref{qme}) (or,
equivalently, in the Wilson coefficients $C_\pm$).
This ambiguity consists in what part of the full 
(non-logarithmic) ${\cal O}(\alpha_s)$ term is attributed to the matrix
elements, and what part to the Wilson coefficients. In other words,
coefficients and matrix elements become {\it scheme dependent},
that is, dependent on the renormalization scheme, beyond the leading
logarithmic approximation. The scheme dependence is 
unphysical and cancels in the product of coefficients and matrix
elements. Of course, both quantities have to be evaluated in the
same scheme to obtain a consistent result. The renormalization scheme
is determined in particular by the subtraction constants
(minimal or non-minimal subtraction of $1/\varepsilon$ poles), and
also by the definition of $\gamma_5$ used in $D\not= 4$ dimensions
in the context of dimensional regularization. 
\item
Finally, the effective Hamiltonian (\ref{hcqpm}) can be considered
as a modern version of the old Fermi theory for weak interactions.  
It is a systematic 
low-energy approximation to the standard model for kaon decays
and provides the basis for any further analysis.
\end{itemize}

\subsection{Renormalization Group}

Let us have a closer look at the Wilson coefficents, which read
explicitly
\begin{equation}\label{cpmg}
C_\pm=1+\frac{\alpha_s(\mu)}{4\pi}\frac{\gamma^{(0)}_\pm}{2}
\ln\frac{\mu^2}{M^2_W}
\qquad
\gamma^{(0)}_\pm =\left\{
\begin{array}{c}
4\\ -8
\end{array}\right.
\end{equation}
where we have now specified the exact form of the
${\cal O}(\alpha_s\,{\rm log})$ correction.
Numerically the factor $\alpha_s(\mu) \gamma^{(0)}_\pm/(8\pi)$
is about $+7\%$ ($-14\%$), a reasonable size for a perturbative
correction (we used $\alpha_s(\mu=1\,{\rm GeV})=0.43$).
However, this term comes with a large logarithmic factor of
$\ln(\mu^2/M^2_W)=-8.8$, for an appropriate scale of $\mu=1\,{\rm GeV}$.
The total correction to $C_\pm=1$ in (\ref{cpmg}) is then
$-60\%$ ($120\%$)! Obviously, the presence of the large logarithm
spoils the validity of a straightforward perturbative expansion,
despite the fact that the coupling constant itself is still
reasonably small. This situation is quite common in renormalizable
quantum field theories. Logarithms appear naturally and can become
very large when the problem involves very different scales.
The general situation is indicated in the following table, where we 
display the form of the correction terms in higher orders, denoting
$\ell\equiv\ln(\mu/M_W)$
\begin{equation}\nonumber
\begin{array}{cccc}
{\rm LL} & {\rm NLL} & & \\
\alpha_s\ell & \alpha_s & & \\
\alpha^2_s\ell^2 & \alpha^2_s\ell & \alpha^2_s & \\
\alpha^3_s\ell^3 & \alpha^3_s\ell^2 & \alpha^3_s\ell & \alpha^3_s \\
\downarrow & \downarrow & & \\
{\cal O}(1) & {\cal O}(\alpha_s) & & \\
\end{array}
\end{equation}
In ordinary perturbation theory the expansion is organized according
to powers of $\alpha_s$ alone, corresponding to the rows in the
above scheme. This approach is invalidated by the large logarithms
since $\alpha_s\ell$, in contrast to $\alpha_s$, is no longer a
small parameter, but a quantity of order 1.
The problem can be resolved by resumming the terms 
$(\alpha_s\ell)^n$ to all orders $n$.
The expansion is then reorganized in terms of columns of the
above table. The first column is of ${\cal O}(1)$ and yields
the leading logarithmic approximation, the second column gives a
correction of relative order $\alpha_s$, and so forth. 
Technically the reorganization is achieved by solving the
renormalization group equation (RGE) for the Wilson coefficients.
The RGE is a differential equation describing the change of 
$C_\pm(\mu)$ under a change of scale. To leading order this equation
can be read off from (\ref{cpmg})
\begin{equation}\label{rgecpm}
\frac{d}{d\ln\mu} C_\pm(\mu)=
\frac{\alpha_s}{4\pi}\gamma^{(0)}_\pm\cdot C_\pm(\mu)
\end{equation}
$(\alpha_s/4\pi)\gamma^{(0)}_\pm$ are called the anomalous dimensions
of $C_\pm$. To understand the term ``dimension'', compare with the
following relation for the quantity $\mu^n$, which has (energy)
dimension $n$:
\begin{equation}\label{dmun}
\frac{d}{d\ln\mu}\mu^n = n\cdot \mu^n
\end{equation}
The analogy is obvious. 
Of course, the $C_\pm(\mu)$ are dimensionless numbers in the
usual sense; they can depend on the energy scale $\mu$ only because
there is another scale, $M_W$, present under the logarithm in 
(\ref{cpmg}). Their ``dimension'' is therefore more precisely
called a scaling dimension, measuring the rate of change of $C_\pm$
with a changing scale $\mu$. The nontrivial scaling dimension
derives from ${\cal O}(\alpha_s)$ loop corrections and is thus a 
genuine quantum effect. Classically the coefficients are 
scale invariant, $C_\pm\equiv 1$. Whenever a symmetry that holds
at the classical level is broken by quantum effects, we speak of an
``anomaly''. Hence, the $\gamma^{(0)}_\pm$ represent the
anomalous (scaling) dimensions of the Wilson coefficients.

We can solve (\ref{rgecpm}), using
\begin{equation}\label{asb0c}
\frac{d\alpha_s}{d\ln\mu}=-2\beta_0\frac{\alpha^2_s}{4\pi}
\quad
\beta_0=\frac{33-2f}{3}\quad C_\pm(M_W)=1
\end{equation}
and find
\begin{equation}\label{cpmll}
C_\pm(\mu)=
\left[\frac{\alpha_s(M_W)}{\alpha_s(\mu)}
  \right]^{\frac{\gamma^{(0)}_\pm}{2\beta_0}}=
\left[\frac{1}{1+\beta_0\frac{\alpha_s(\mu)}{4\pi}\ln\frac{M^2_W}{\mu^2}}
\right]^{\frac{\gamma^{(0)}_\pm}{2\beta_0}}
\end{equation}
This is the solution for the Wilson coefficients $C_\pm$ in leading
logarithmic approximation, that is to leading order in RG improved
perturbation theory. The all-orders resummation of $\alpha_s\,{\rm log}$
terms is apparent in the final expression in (\ref{cpmll}).

\subsection{$\Delta I=1/2$ Rule}

At this point, and before continuing with the construction of
the complete $\Delta S=1$ Hamiltonian, it is interesting to
discuss a first application of the results we have derived so far.

\begin{figure}[t]
\hspace*{3.2cm}\epsfig{figure=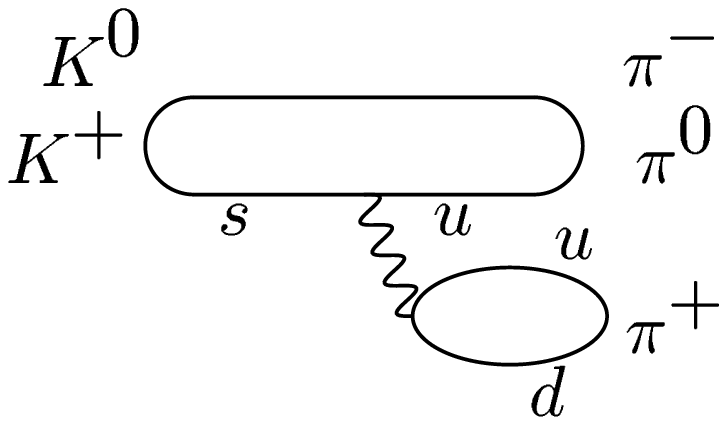,width=7cm,height=5cm}
\vspace*{-2cm}
\caption{$K^0\to\pi^+\pi^-$ and $K^+\to\pi^+\pi^0$. \label{fig:kp0pp}}
\end{figure}

Let us consider the weak decays into two pions of a neutral
kaon, $K_S\to\pi^+\pi^-$, and a charged kaon, $K^+\to\pi^+\pi^0$,
which are sketched in Fig. \ref{fig:kp0pp}. 
The two cases look
very much the same, except that the spectator quark is a $u$ quark
for the charged kaon and a $d$ quark for the neutral one.
Naively one would therefore expect very similar decay rates.
The experimental facts are, however, strikingly different:
\begin{equation}\label{di12}
\frac{\Gamma(K_S\to\pi^+\pi^-)}{\Gamma(K^+\to\pi^+\pi^0)}
\approx 450\approx (21.2)^2
\end{equation}
To get a hint as to where this huge difference in the decay rates may
come from we have to analyze the isospin structure of the decays.
A kaon state has isospin $I=1/2$. Taking into account Bose symmetry,
one finds that two pions from the decay of a $K$ meson can only be
in a state of isospin $0$ and $2$. More specifically,
$|\pi^+\pi^-\rangle$ has both $I=0$ and $I=2$ components,
while $|\pi^+\pi^0\rangle$ is a pure $I=2$ state.
The change in isospin is then as follows
\begin{eqnarray}
K^+\to\pi^+\pi^0  &&  \Delta I=3/2 \\ 
K^0\to\pi^+\pi^-  &&  \Delta I=1/2,\, 3/2
\end{eqnarray}
In particular, $K^+\to\pi^+\pi^0$ is a pure $\Delta I=3/2$
transition. The large ratio in (\ref{di12}) means that $\Delta I=1/2$
transitions are strongly enhanced. This empirical feature is refered
to as the $\Delta I=1/2$ rule.

We next take a closer look at the isospin properties of the
effective Hamiltonian. Using the Fierz identities of the Dirac
matrices the operators $Q_\pm$ can be rewritten as
\begin{eqnarray}\label{qpmfz}
Q_\pm &=& 
(\bar s_i u_i)_{V-A}(\bar u_j d_j)_{V-A}\pm
(\bar s_i u_j)_{V-A}(\bar u_j d_i)_{V-A}= \nonumber \\
&=&
(\bar s_i u_i)_{V-A}(\bar u_j d_j)_{V-A}\pm
(\bar s_i d_i)_{V-A}(\bar u_j u_j)_{V-A}
\end{eqnarray}
where now all quark bilinears appear uniformly as colour singlets.
Retaining only the flavour structure, but dropping colour and
Dirac labels for ease of notation, the Hamiltonian (\ref{hcqpm})
has the form
\begin{eqnarray}\label{heffiso}
{\cal H}_{eff} &\sim&
\left[\frac{\alpha_s(M_W)}{\alpha_s(\mu)}\right]^{6/25}
\left( (\bar su)(\bar ud)+(\bar sd)(\bar uu)\right) + \nonumber \\  
&+&
\left[\frac{\alpha_s(M_W)}{\alpha_s(\mu)}\right]^{-12/25}
\left( (\bar su)(\bar ud)-(\bar sd)(\bar uu)\right) 
\end{eqnarray}
We can now see that the operator $Q_-$, in the second line of
(\ref{heffiso}), is a pure $\Delta I=1/2$ operator:
$u$ and $d$ appear in the combination
\begin{equation}
ud-du {\hat =} |\uparrow\downarrow\rangle -|\downarrow\uparrow\rangle
\end{equation}
which has isospin 0.
The strange quark is also an isospin singlet. The isospin of
$Q_-$ is therefore determined by the factor $\bar u$, which has
isospin $1/2$.

{}From the Wilson coefficients we have calculated we observe that
the contribution from $Q_-$ receives a relative enhancement over
$Q_+$ in (\ref{heffiso}) by a factor
\begin{equation}\label{cmcp}
\left[\frac{\alpha_s(\mu)}{\alpha_s(M_W)}\right]^{18/25}
\approx\left\{
\begin{array}{ll}
2.6 & \ \ \ \mu=1\,{\rm GeV}\\
3.4 & \ \ \ \mu=0.6\,{\rm GeV}
\end{array}\right.
\end{equation}
Qualitatively, this is precisely what we need: $Q_-$,
which is purely $\Delta I=1/2$ and can thus only contribute to
$K^0\to\pi^+\pi^-$, but not to $K^+\to\pi^+\pi^0$, is re-inforced
by the short-distance QCD dynamics.
Quantitatively, however, the RG improved QCD effect falls still
short of explaining the amplitude ratio $21.2$ in (\ref{di12})
by a sizable factor. We might be tempted to decrease $\mu$, which
enhances the effect, but we are not allowed to go much below
$\mu=1\,{\rm GeV}$ where perturbation theory would cease to be valid.
The remaining enhancement has to come from nonperturbative
contributions in the matrix elements.
Nevertheless it is interesting to see how already the short-distance
QCD corrections provide the first step towards a dynamical
explanation of the $\Delta I=1/2$ rule.

\subsection{$\Delta S=1$ Effective Hamiltonian}

\begin{figure}[h]
\hspace*{4.2cm}\epsfig{figure=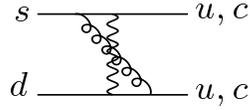,width=7cm,height=5cm}
\vspace*{-3cm}
\caption{QCD correction to $W$ exchange. \label{fig:s1cc}}
\end{figure}

\begin{figure}[h]
\hspace*{4.2cm}\epsfig{figure=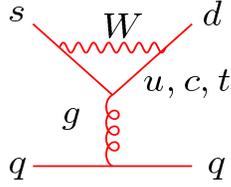,width=7cm,height=5cm}
\vspace*{-2.5cm}
\caption{QCD-penguin diagram. \label{fig:s1peng}}
\end{figure}

In this section we will complete the discussion of the
$\Delta S=1$ effective Hamiltonian.
So far we have considered the operators
\begin{eqnarray}\label{q12def}
Q_1 &=& (\bar s_i u_j)_{V-A}(\bar u_j d_i)_{V-A} \\
Q_2 &=& (\bar s_i u_i)_{V-A}(\bar u_j d_j)_{V-A}
\end{eqnarray}
which come from the simple $W$-exchange graph and the corresponding
QCD corrections (Fig. \ref{fig:s1cc}).
In addition, there is a further type of diagram at ${\cal O}(\alpha_s)$,
which we have omitted until now: the QCD-penguin diagram shown
in Fig. \ref{fig:s1peng}. 
It gives rise to four new operators
\begin{eqnarray}\label{q36def}
Q_3 &=& (\bar s_i d_i)_{V-A}\sum_q(\bar q_j q_j)_{V-A} \\
Q_4 &=& (\bar s_i d_j)_{V-A}\sum_q(\bar q_j q_i)_{V-A} \\
Q_5 &=& (\bar s_i d_i)_{V-A}\sum_q(\bar q_j q_j)_{V+A} \\
Q_6 &=& (\bar s_i d_j)_{V-A}\sum_q(\bar q_j q_i)_{V+A} 
\end{eqnarray}
Two structures appear when the light-quark current $(\bar qq)_V$
from the bottom end of the diagram is split into $V-A$ and $V+A$
parts. In turn, each of those comes in two colour forms in a way
similar to $Q_1$ and $Q_2$.

The operators $Q_1,\ldots ,Q_6$ mix under renormalization, that is
the RGE for their Wilson coefficients is governed by a matrix
of anomalous dimensions, generalizing (\ref{rgecpm}).
In this way the RG evolution of $C_{1,2}$ affects the evolution
of $C_3,\ldots , C_6$. On the other hand $C_{1,2}$ remain unchanged
in the presence of the penguin operators $Q_3,\ldots ,Q_6$,
so that the results for $C_{1,2}$ derived above are still valid.

\begin{figure}[h]
\hspace*{4.2cm}\epsfig{figure=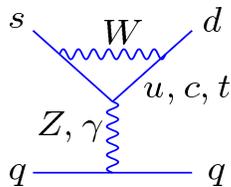,width=7cm,height=5cm}
\vspace*{-2.5cm}
\caption{Electroweak penguin. \label{fig:s1ewp}}
\end{figure}

For some applications (e.g. $\varepsilon'/\varepsilon$) higher order
electroweak effects need to be taken into account. They arise from
$\gamma$- or $Z$-penguin diagrams (Fig. \ref{fig:s1ewp})
and also from $W$-box diagrams. Four additional operators arise
from this source. They have a form similar to the QCD penguins,
but a different isospin structure, and read 
($e_q$ are the quark charges, $e_{u,c}=+2/3$, $e_{d,s,b}=-1/3$)
\begin{eqnarray}\label{q710def}
Q_7 &=& \frac{3}{2}(\bar s_i d_i)_{V-A}\sum_q e_q(\bar q_j q_j)_{V+A} \\
Q_8 &=& \frac{3}{2}(\bar s_i d_j)_{V-A}\sum_q e_q(\bar q_j q_i)_{V+A} \\
Q_9 &=& \frac{3}{2}(\bar s_i d_i)_{V-A}\sum_q e_q(\bar q_j q_j)_{V-A} \\
Q_{10} &=& \frac{3}{2}(\bar s_i d_j)_{V-A}\sum_q e_q(\bar q_j q_i)_{V-A} 
\end{eqnarray}

The construction of the effective Hamiltonian follows the principles
we have discussed in the previous sections.
First the Wilson coefficients $C_i(\mu_W)$, $i=1,\ldots, 10$,
are determined at a large scale $\mu_W={\cal O}(M_W,m_t)$ to a given
order in perturbation theory. In this step both the $W$ boson and the
heavy top quark are integrated out. Since the renormalization scale
is chosen to be $\mu_W={\cal O}(M_W,m_t)$, no large logarithms
appear and straightforward perturbation theory can be used for the
matching calculation.
The anomalous dimensions are computed from the divergent parts of
the operator matrix elements, which correspond to the 
UV-renormalization of the Wilson coefficients. Solving the
RGE the $C_i$ are evolved from $\mu_W$ to a scale 
$\mu_b={\cal O}(m_b)$ in a theory with $f=5$ active flavours
$q=u,d,s,c,b$. At this point the $b$ quark (which can appear in loops)
is integrated out by calculating the matching conditions from a 
five-flavour to a four-flavour theory, where only $q=u,d,s,c$
are active. This procedure is repeated by integrating out
charm at $\mu_c={\cal O}(m_c)$ and matching onto an $f=3$
flavour theory.
One finally obtains the coefficients $C_i(\mu)$ at a scale
$\mu <\mu_c$, describing an effective theory where only
$q=u,d,s$ (and gluons of course) are active degrees of freedom.
The terms taken into account in the RG improved perturbative
evaluation of $C_i(\mu)$ are, schematically:

\begin{center}
LO: $\left(\alpha_s\ln\frac{M_W}{\mu}\right)^n$,
$\alpha\ln\frac{M_W}{\mu}\left(\alpha_s\ln\frac{M_W}{\mu}\right)^n$

NLO: $\alpha_s\left(\alpha_s\ln\frac{M_W}{\mu}\right)^n$,
$\alpha\left(\alpha_s\ln\frac{M_W}{\mu}\right)^n$
\end{center}

\noindent
at leading and next-to-leading order, respectively.
Here $\alpha$ is the QED coupling, refering to the electroweak
corrections.

The final result for the $\Delta S=1$ effective Hamiltonian
(with 3 active flavours) can be written as
\begin{equation}\label{hds1}
{\cal H}^{\Delta S=1}_{eff}=\frac{G_F}{\sqrt{2}}\lambda_u
\sum^{10}_{i=1}\left(z_i(\mu)-\frac{\lambda_t}{\lambda_u}y_i(\mu)
\right)Q_i\, + \,{\rm h.c.}
\end{equation}
where $\lambda_p\equiv V^*_{ps}V_{pd}$.
In principle there are three different CKM factors,
$\lambda_u$, $\lambda_c$ and $\lambda_t$, corresponding to the
different flavours of up-type quarks that can participate
in the charged-current weak interaction. Using CKM unitarity, one
of them can be eliminated. If we eliminate $\lambda_c$,
we arrive at the CKM structure of (\ref{hds1}).

The Hamiltonian in (\ref{hds1}) is the basis for computing
nonleptonic kaon decays within the standard model, in particular
for the analysis of direct CP violation.
When new physics is present at some higher energy scale,
the effective Hamiltonian can be derived in an analogous way.
The matching calculation at the high scale $\mu_W$ will give
new contributions to the coefficients $C_i(\mu_W)$, the initial
conditions for the RG evolution. In general, new operators may
also be induced.
The Wilson coefficients $z_i$ and $y_i$ are known in the standard
model at NLO. A more detailed account of 
${\cal H}^{\Delta S=1}_{eff}$ and information on the technical
aspects of the necessary calculations can be found in
\cite{BBL} and \cite{AJB1}.

\subsection{Chiral Perturbation Theory}

An additional tool for kaon physics, complementary to the
OPE-based effective Hamiltonian formalism, is
chiral perturbation theory ($\chi$PT).
The present section gives a brief and elementary introduction
into this subject. For other, more detailed discussions
we refer the reader to \cite{DI,HG,AP2,CI}.

\subsubsection{Preliminaries}

The QCD Lagrangian for three light flavours $q=(u,d,s)^T$ can be
written in terms of left-handed and right-handed fields,
$q_{L,R}=(1\mp\gamma_5) q/2$, in the form
\begin{equation}\label{lqcd}
{\cal L}_{QCD}=\bar q_L i\not\!{\cal D}q_L+\bar q_R i\not\!{\cal D}q_R
-\bar q_L{\cal M}q_R-\bar q_R{\cal M}^\dagger q_L
\end{equation}
where ${\cal M}={\rm diag}(m_u,m_d,m_s)$.
If ${\cal M}$ is put to zero, ${\cal L}_{QCD}$ is invariant under
a global $SU(3)_L\otimes SU(3)_R$ symmetry
\begin{eqnarray}\label{su3lr}
q_L &\to&  L q_L \\
q_R &\to& R q_R
\end{eqnarray}
with $L$ and $R$ (independent) $SU(3)$ transformations.
The explicit breaking of this {\it chiral symmetry\/} through a
nonzero ${\cal M}$ is a small effect and can be treated as a
perturbation. Simultaneously, chiral symmetry is not reflected in
the hadronic spectrum, so it must also be spontaneously broken by
the dynamics of QCD. For instance the octet of light pseudoscalar
mesons
\begin{equation}\label{phi33}
\Phi\equiv T^a\pi^a=
\left(\begin{array}{ccc}
\frac{\pi^0}{\sqrt{2}}+\frac{\eta}{\sqrt{6}} & \pi^+ & K^+ \\
\pi^- & -\frac{\pi^0}{\sqrt{2}}+\frac{\eta}{\sqrt{6}} & K^0 \\
K^- & \bar K^0 & -\frac{2\eta}{\sqrt{6}}
\end{array}\right)
\end{equation}
is not accompanied in the spectrum of hadrons by a
similar octet of mesons with opposite parity and comparable mass.
On the other hand the octet $\Phi$, comprising the lightest existing
hadrons, is the natural candidate for the octet of Goldstone bosons
expected from the pattern of spontaneous chiral symmetry breaking
\begin{equation}
SU(3)_L\otimes SU(3)_R \to SU(3)
\end{equation}
down to group of ordinary flavour $SU(3)$, 
where $q_{L,R}\to U q_{L,R}$. 

The mesons in $\Phi$ are not strictly massless due to the explicit
breaking of chiral symmetry caused by ${\cal M}$ and are thus often
refered to as pseudo-Goldstone bosons. Still they are the lightest
hadrons and they are separated by a mass gap from the higher excitations
of the light-hadron spectrum. (The masses of the latter remain of
order $\Lambda_{QCD}$, while the masses of $\Phi$ vanish in the
limit ${\cal M}\to 0$.)

The idea of $\chi$PT is to write a low-energy effective theory where
the only dynamical degrees of freedom are the eight pseudo-Goldstone
bosons.
This is appropriate for low-energy interactions where the higher states
are not kinematically accessible. Their virtual presence will however
be contained in the coupling constants of $\chi$PT.
The guiding principles for the construction of  $\chi$PT are the
chiral symmetry of QCD and an expansion in powers of momenta and
quark masses. By constructing the most general Lagrangian for
$\Phi$ compatible with the symmetries of QCD, the framework
is {\it model independent}. By restricting the accuracy to a given order
in the momentum expansion, only a finite number of terms are possible
and the framework becomes also {\it predictive}.
A finite number of couplings needs to be fixed from experiment; once
this is done, predictions can be made. $\chi$PT is a nonperturbative
approach as it does not rely on any expansion in the QCD
coupling $\alpha_s$.

Both $\chi$PT and the quark-level effective Hamiltonian
${\cal H}_{eff}$ are low-energy effective theories applicable to
kaon decays. What is the difference and how are these two
approaches related?
The essential, and obvious, difference is that $\chi$PT is
formulated directly in terms of hadrons, ${\cal H}_{eff}$ in terms
of quarks and gluons.
The advantage of $\chi$PT is therefore the direct applicability
to physical, hadronic amplitudes, without the need to deal with
the complicated hadronic matrix elements of quark-level operators.
The advantage of ${\cal H}_{eff}$, on the other hand, is the direct
link to short-distance physics, which is encoded in the Wilson
coefficients. This type of information is important in the context
of CP violation or in the search for new physics.
In $\chi$PT such information is hidden in the coupling constants,
which are not readily calculable and need to be fixed experimentally.
{}From these considerations it is clear that  ${\cal H}_{eff}$ is more
useful for applications where short-distance physics is essential
(CP violation, $\varepsilon'/\varepsilon$), whereas $\chi$PT
is especially suited to deal with long-distance dominated
quantities, which are hard to come by otherwise.
To relate the two descriptions directly is not an easy task and
has so far not been accomplished.
A calculation of the couplings of $\chi$PT from the quark picture,
establishing a link between ${\cal H}_{eff}$ and $\chi$PT, requires one to
solve QCD nonperturbatively, which is not possible at present.

\subsubsection{$SU(3)$ Transformations}

Before describing the explicit construction of $\chi$PT it will
be useful to recall a few important properties of $SU(3)$
transformations and to introduce some convenient notation.
We define
\begin{equation}\label{quds}
(q^1,q^2,q^3)\equiv (u,d,s) \qquad
(q_1,q_2,q_3)\equiv (\bar u,\bar d,\bar s)
\end{equation}
and by $U^i_{\ j}$ the components of a generic $SU(3)$ matrix $U$.
By definition, changing upper into lower indices, and vice versa,
corresponds to complex conjugation, thus
\begin{equation}\label{usdij}
U^{\ j}_i\equiv U^{* i}_{\ \ j}=U^{\dagger j}_{\ \ i}
\end{equation} 
The unitarity of $U$ implies
\begin{equation}\label{uudel1}
U^{\ i}_k U^k_{\ j}=U^{\dagger i}_{\ \ k} U^k_{\ j}=\delta^i_{\ j}
\end{equation}
\begin{equation}\label{uudel2}
U^i_{\ k} U^{\ k}_j=U^i_{\ k} U^{\dagger k}_{\ \ j}=\delta^i_{\ j}
\end{equation}
Also,
\begin{equation}\label{detueps}
{\rm det}\, U=\varepsilon^{ijk} U^1_{\ i}U^2_{\ j}U^3_{\ k} = 1
\end{equation}
The fundamental $SU(3)$ triplet $q^i$ and anti-triplet $q_i$
transform, respectively, as
\begin{equation}\label{qij}
q^i\to U^i_{\ j} q^j
\end{equation}
\begin{equation}\label{qijb}
q_i\to U^{\ j}_i q_j
\end{equation}
It follows from the above that the singlet $q^k q_k$ as well as the
Kronecker symbol $\delta^i_{\ j}$ and the totally antisymmetric
tensor $\varepsilon^{ijk}$ are invariant under $SU(3)$ transformations.

Higher dimensional representations can also be built.
For example, the traceless tensor
\begin{equation}\label{sij}
S^i_{\ j}=q^i q_j-\frac{1}{3}\delta^i_{\ j}\, q^k q_k
\end{equation}
is an irreducible representation of $SU(3)$. Its eight components
constitute an $SU(3)$ octet, which transforms as
\begin{equation}\label{stra}
S^i_{\ j}\to U^i_{\ k} U^{\ l}_j S^k_{\ l}
\end{equation}
We next define the objects
\begin{equation}\label{rij}
r^i=\varepsilon^{ijk} q_j q_k
\end{equation}
\begin{equation}\label{rijb}
r_i=\varepsilon_{ijk} q^j q^k
\end{equation}
They transform in the same way as the fundamental
triplet and anti-triplet in (\ref{qij}) and (\ref{qijb}),
respectively. We show this for (\ref{rij}). 
{}From (\ref{qijb}) and
using (\ref{usdij}) and (\ref{uudel2}), we have
\begin{eqnarray}\label{ritra}
r^i &=& \varepsilon^{ijk} q_j q_k \to \nonumber \\
&& \varepsilon^{ijk}U^{\ l}_j U^{\ m}_k q_l q_m =
U^i_{\ s}\,\varepsilon^{njk}U^{\ s}_n U^{\ l}_j U^{\ m}_k q_l q_m=
\nonumber \\
&& U^i_{\ s}\,\varepsilon^{njk}
U^{\dagger s}_{\ \ n} U^{\dagger l}_{\ \ j} U^{\dagger m}_{\ \ \, k}\, q_l q_m=
U^i_{\ s}\,\varepsilon^{slm} ({\rm det}U^\dagger) q_l q_m=U^i_{\ s} r^s
\end{eqnarray}
which proves our assertion.

Let us consider two simple applications of this formalism.

a) The meson field $\Phi$ in (\ref{phi33}) corresponds to the
quark-level tensor $S^i_{\ j}$ and both transform as octets under
$SU(3)$. The connection can be seen by writing out the components
of $S^i_{\ j}$ and comparing with (\ref{phi33}). One recovers
the quark flavour composition of the meson states. For example
\begin{equation}\label{p12}
\Phi^1_{\ 2}=\pi^+\, \hat =\, S^1_{\ 2}=q^1 q_2 = u\bar d
\end{equation}
\begin{equation}\label{p33}
\Phi^3_{\ 3}=-\frac{2}{\sqrt{6}}\eta\, \hat =\, S^3_{\ 3}=
q^3 q_3-\frac{1}{3} q^k q_k=-\frac{1}{3}(u\bar u+d\bar d-2 s\bar s)
\end{equation}
The transformation law for $\Phi^i_{\ j}$ is the same as for 
$S^i_{\ j}$ in (\ref{stra}) or, in matrix notation,
\begin{equation}\label{phiuu}
\Phi\to U\Phi U^\dagger
\end{equation}

b) In section 3.3 we have seen from the discussion of the
$\Delta I=1/2$ rule that the largely dominant part of the weak
Hamiltonian is contributed by the pure $\Delta I=1/2$ operator
$Q_-$. We will now show the important property that $Q_-$ transforms
as the component of an octet under $SU(3)_L$.
To see this we first note that the operator
\begin{equation}\label{qrij}
Q^i_{\ j}=r^i r_j-\frac{1}{3}\delta^i_{\ j}\, r^k r_k
\end{equation}
is an $SU(3)$ octet. This follows because the $r^i$ in (\ref{rij}),
(\ref{rijb}) transform as the $q^i$, and $S^i_{\ j}$ in (\ref{sij})
is an octet. We next show that the (2,3) component of $Q^i_{\ j}$
indeed has the flavour structure of $Q_-$:

\begin{eqnarray}\label{q23qm}
Q^2_{\ 3} &=& r^2 r_3=(q_3 q_1-q_1 q_3)(q^1 q^2-q^2 q^1)\nonumber\\
&{\hat =}& \, (\bar su)(\bar ud)-(\bar sd)(\bar uu)-
(\bar uu)(\bar sd)+(\bar ud)(\bar su) \,\hat =\, 2 Q_-
\end{eqnarray}
Since the quark fields in $Q_-$ are all left-handed, we see that
$Q_-$ transforms as the (2,3) component of an octet $Q^i_{\ j}$
under $SU(3)_L$. Trivially, it is also a singlet under $SU(3)_R$.
Hence, $Q_-$ transforms as a component of a 
$(8_L,1_R)$ under $SU(3)_L\otimes  SU(3)_R$.

The transformation law for $Q^i_{\ j}$ in matrix notation is,
with $L\in SU(3)_L$,
\begin{equation}\label{qlql}
Q\to L Q L^\dagger
\end{equation}
Including the hermitian conjugate we may write
\begin{equation}\label{trl6q}
2(Q_-+Q^\dagger_-)\,\hat =\, Q^2_{\ 3}+Q^3_{\ 2}=
{\rm tr}\, \lambda_6 Q
\end{equation}
where the trace with the Gell-Mann matrix $\lambda_6$ is used
to project out the proper components ($\lambda_6$ is the matrix
with entry 1 at positions (2,3) and (3,2), and 0 otherwise).

It is not hard to see that the penguin operators $Q_3,\ldots, Q_6$
also transform as part of an $(8_L,1_R)$, in the same way as $Q_-$.

\subsubsection{Chiral Lagrangian}

We will now construct explicitly the leading terms of the
chiral Lagrangian. Since we have to write down the most general
form for this Lagrangian to any given order in the momentum expansion,
the specific manner in which chiral symmetry is realized does 
not matter. The most convenient and standard choice is a nonlinear
realization where one introduces the unitary matrix
\begin{equation}\label{sigphi}
\Sigma =\exp\left(\frac{2i}{f}\Phi\right)
\end{equation}
as the basic meson field. Here $f$ is the generic decay constant
for the light pseudoscalars (we have used a normalization in which
$f_\pi=131\,{\rm MeV}$).

The field $\Sigma$ is taken to transform under 
$SU(3)_L\otimes SU(3)_R$ as
\begin{equation}\label{siglr}
\Sigma \to L\Sigma R^\dagger
\end{equation}
with $L\in SU(3)_L$ and $R\in SU(3)_R$. In general, (\ref{siglr})
implies a complicated, nonlinear transformation law for the field
$\Phi$. However, for the special case of an ordinary $SU(3)$
transformation, where $L=R\equiv U$, (\ref{siglr}) becomes equivalent
to (\ref{phiuu}). We thus recover the correct transformation for the
octet $\Phi$ under ordinary $SU(3)$.
We can also see that the vacuum state, which corresponds to
$\Phi\to 0$, hence $\Sigma\to 1$, is invariant under ordinary $SU(3)$
($1\to U\, 1\, U^\dagger=1$) as it must be.
On the other hand, the vacuum is not invariant under the
chiral transformation in (\ref{siglr}):
$1\to L\, 1\, R^\dagger \not= 1$. This corresponds to the
spontaneous breaking of chiral symmetry.
The field (\ref{sigphi}) with the transformation (\ref{siglr})
therefore has the desired properties to describe the pseudo-Goldstone
bosons.

The chiral Lagrangian is constructed as a series in powers of
momenta, or equivalently numbers of derivatives
\begin{equation}\label{lqcdchi}
{\cal L}^{QCD}={\cal L}^{QCD}_2 +  {\cal L}^{QCD}_4 +\ldots
\end{equation}
\begin{equation}\label{lds1chi}
{\cal L}^{\Delta S=1}=
{\cal L}^{\Delta S=1}_2 +  {\cal L}^{\Delta S=1}_4 +\ldots
\end{equation}
${\cal L}^{QCD}$ describes strong interactions, 
${\cal L}^{\Delta S=1}$ $\Delta S=1$ weak interactions, and the
subscripts on the right-hand side indicate the number of derivatives.

The lowest order strong interaction Lagrangian has the form
\begin{equation}\label{lqcd2}
{\cal L}^{QCD}_2=\frac{f^2}{8}\,{\rm tr}
\left[ D_\mu\Sigma D^\mu\Sigma^\dagger +
2 B_0({\cal M}\Sigma^\dagger +\Sigma {\cal M}^\dagger)\right]
\end{equation}
We have written $D_\mu$ for the derivative, which we later will
generalize to a covariant derivative to include electromagnetism.
For the moment we may consider $D_\mu$ as an ordinary derivative.

The terms in (\ref{lqcd2}) have to be built to respect the symmetries
of the QCD Lagrangian in (\ref{lqcd}). For ${\cal M}=0$,
(\ref{lqcd}) is chirally invariant and we are thus looking for
invariants constructed from $\Sigma$ in (\ref{siglr}).
Only trivial terms are possible with zero derivatives, for example
${\rm tr}(\Sigma\Sigma^\dagger)={\rm const.}$
The leading term comes with two derivatives, as anticipated.
The only possible form is
${\rm tr}(D_\mu\Sigma D^\mu\Sigma^\dagger)$. Here
$D_\mu\Sigma D^\mu\Sigma^\dagger$ transforms as $(8_L,1_R)$
\begin{equation}\label{dsdstra}
D_\mu\Sigma D^\mu\Sigma^\dagger \to
L\, D_\mu\Sigma D^\mu\Sigma^\dagger\, L^\dagger
\end{equation}
and taking the trace gives an invariant. Another possibility
would seem to be ${\rm tr}(D^2\Sigma \Sigma^\dagger)$,
but this term differs from
${\rm tr}(D_\mu\Sigma D^\mu\Sigma^\dagger)$ only by a total
derivative.

The second term in (\ref{lqcd2}) breaks chiral symmetry. Its
form can be found by noting that the symmetry breaking term
proportional to ${\cal M}$ in (\ref{lqcd}) would be invariant if
${\cal M}$ was interpreted as an auxiliary field transforming
as ${\cal M}\to L{\cal M}R^\dagger$.
To first order in ${\cal M}$ and to lowest order in derivatives
this leads to the mass term in (\ref{lqcd2}).
For ${\cal M}\to L{\cal M}R^\dagger$ it would be invariant.
For ${\cal M}$ fixed to the diagonal mass matrix it breaks
chiral symmetry in the appropriate way. We will soon find that this
term indeed counts as two powers of momentum.
The second order Lagrangian ${\cal L}^{QCD}_2$ is then complete.
The factor in front of the first term in (\ref{lqcd2}) is fixed
by the requirement that the kinetic term for the mesons be normalized
in the canonical way. There are no additional parameters for this
contribution. The second term in (\ref{lqcd2}) comes with a coupling
$B_0$. This new parameter is related to the meson masses. To see this 
more clearly, we can extract the kinetic terms from (\ref{lqcd2}) by
expanding to second order in the field $\Phi$. If we keep, for example,
only the contributions with neutral kaons $K$, $\bar K$, we find
(up to an irrelevant additive constant)
\begin{equation}\label{lkkin}
{\cal L}^{QCD}_{2,K^0 kin}=\partial^\mu\bar K\partial_\mu K
   -B_0(m_s+m_d) \bar K K
\end{equation}
{}From (\ref{lkkin}) and similar relations for the other mesons we
obtain expressions for the pseudo-Goldstone boson masses in terms
of the quark masses and the parameter $B_0={\cal O}(\Lambda_{QCD})$ 
\begin{eqnarray}\label{m2bmq}
m^2_{K^0} &=& B_0(m_s+m_d) \nonumber\\
m^2_{K^+} &=& B_0(m_s+m_u) \\
m^2_{\pi^+} &=& B_0(m_u+m_d) \nonumber
\end{eqnarray}
The meson masses squared are proportional to linear combinations
of quark masses. This also clarifies why one factor of ${\cal M}$
is equivalent to two powers of momenta in the usual chiral
counting (see (\ref{lqcd2})).

Expanding (\ref{lqcd2}) beyond second order in $\Phi$ we obtain
terms describing strong interactions among the mesons, such as
$\pi$-$\pi$ scattering.

We next need to determine the form of ${\cal L}^{\Delta S=1}_2$.
As we have seen in sec. 3.3, the dominant contribution to the 
weak Hamiltonian comes from the component of an $(8_L,1_R)$ operator 
as shown in (\ref{trl6q}). Since we know empirically from the
$\Delta I=1/2$ rule that the enhancement of this piece of the
weak Hamiltonian is quite strong, we shall here make the additional
approximation to keep only this contribution and drop the rest
(related to the operator $Q_+$). This is a reasonable approximation
for many applications.
With the results we have derived so far, it is then easy to write down
the correct form for ${\cal L}^{\Delta S=1}_2$.
The structure with two derivatives and the correct transformation
properties as an $(8_L,1_R)$ is given in (\ref{dsdstra}). According
to (\ref{trl6q}) we simply need to take the trace with $\lambda_6$
to obtain the right components. Factoring out basic weak interaction
parameters we can write
\begin{equation}\label{lds12}
{\cal L}^{\Delta S=1}_2=\frac{G_F}{\sqrt{2}}|V^*_{us}V_{ud}|g_8
\frac{f^4}{4}\, {\rm tr}\, \lambda_6 D_\mu\Sigma D^\mu\Sigma^\dagger
\end{equation}
This Lagrangian introduces one additional parameter, the octet
coupling $g_8$.
Eq. (\ref{lds12}) already contains the usual hermitian conjugate
part (see (\ref{trl6q})). We have neglected the small CP violating
effects and factored out the common leading CKM term
$V^*_{us}V_{ud}=V_{us}V^*_{ud}=|V^*_{us}V_{ud}|$.

\subsubsection{$K_S\to\pi^+\pi^-$ from ${\cal L}^{\Delta S=1}_2$}

In order to make predictions, we first need to fix the constant
$g_8$. We can use the dominant nonleptonic decay $K_S\to\pi^+\pi^-$
for this purpose. Expanding the interaction term in (\ref{lds12})
to third order in $\Phi$, and keeping only $K$, $\bar K$, $\pi^+$
and $\pi^-$, we find
\begin{equation}\label{l6dsds}
{\rm tr}\, \lambda_6 \partial\Sigma\partial\Sigma^\dagger=
-\frac{4i}{f^3}\left[\pi^+\partial\bar K\partial\pi^- -
\partial K\pi^-\partial\pi^+ +(K-\bar K)\partial\pi^+\partial\pi^-
\right]
\end{equation}
Neglecting CP violation we have $K_1\equiv K_S$, $K_2\equiv K_L$,
where $CP\, K_{1,2}=\pm K_{1,2}$, and ($CP\, K=-\bar K$)
\begin{equation}\label{k12kkbar}
K_{1,2}=\frac{K\mp\bar K}{\sqrt{2}}
\end{equation}
Expressing $K$, $\bar K$ in terms of $K_{1,2}$ we obtain for
the square brackets in (\ref{l6dsds})
\begin{equation}\label{k1k2pp}
\left[\ldots\right]=
\frac{1}{\sqrt{2}}\pi^+\partial\pi^-(\partial K_2-\partial K_1)-
\frac{1}{\sqrt{2}}\pi^-\partial\pi^+(\partial K_2+\partial K_1)+
\sqrt{2}K_1\partial\pi^+\partial\pi^-
\end{equation}
{}From (\ref{k1k2pp}) the Feynman amplitudes for $K_{1,2}\to\pi^+\pi^-$
can be read off. Denoting the momentum of the kaon, $\pi^+$, $\pi^-$
by $k$, $p_1$, $p_2$, respectively, we get for $K_1$
\begin{equation}\label{brak1}
\left[\ldots\right]_{K_1}\to
-\frac{1}{\sqrt{2}}\left( 2 p_1\cdot p_2+k\cdot(p_1+p_2)\right)
=-\sqrt{2}(m^2_K-m^2_\pi)
\end{equation}
One may check that the corresponding amplitude for $K_2\to\pi^+\pi^-$
gives zero, as required by CP symmetry.

{}From (\ref{brak1}) and (\ref{lds12}) we obtain the decay amplitude
\begin{equation}
A(K_1\to\pi^+\pi^-)=i G_F g_8 |\lambda_u| f_\pi (m^2_K-m^2_\pi)
\end{equation}
This gives the branching ratio
\begin{equation}\label{bkspipi}
B(K_S\to\pi^+\pi^-)=\tau_{K_S}\frac{\sqrt{m^2_K-4m^2_\pi}}{16\pi m^2_K}
\left(m^2_K-m^2_\pi\right)^2 G^2_F g^2_8 |\lambda_u|^2 f^2_\pi
\end{equation}
Using
\begin{equation}
\tau_{K_S}=1.3573\cdot 10^{14}\,{\rm GeV}^{-1}\qquad
B(K_S\to\pi^+\pi^-)=0.6861
\end{equation}
we find
\begin{equation}\label{g8num}
g_8\simeq 5.2
\end{equation}
We have thus determined $g_8$ to lowest order in $\chi$PT.
Using this result, we can make predictions. For instance,
expanding (\ref{lds12}) to fourth order in $\Phi$, we can derive
amplitudes for the decays $K\to 3\pi$.
In this manner $\chi$PT relates processes with different numbers of
soft pions. Such relations, also known as the soft-pion theorems
of current algebra, are nicely summarized in the framework of $\chi$PT
in terms of the lowest-order chiral Lagrangians.
Other important applications are radiative decays as 
$K_S\to\gamma\gamma$, to which we will come back in the following
paragraph.

So far we have worked at tree level, which is sufficient at 
${\cal O}(p^2)$. At the next order, ${\cal O}(p^4)$, one has to consider
both tree-level contributions of the ${\cal O}(p^4)$ terms in the
Lagrangians (\ref{lqcdchi}), (\ref{lds1chi}), and one-loop diagrams
with interactions from the ${\cal O}(p^2)$ Lagrangians.
The loop diagrams are in general divergent. The divergences are
absorbed by renormalizing the couplings at ${\cal O}(p^4)$.
An example will be described below.

\subsubsection{Radiative $K$ Decays}

Electromagnetism and the photon field $A_\mu$ can be included
in $\chi$PT in the usual way.
The $U(1)$ gauge transformation for the meson fields is
\begin{equation}\label{phiu1}
\Sigma'=U\Sigma U^\dagger\Rightarrow
\Phi'=U\Phi U^\dagger,\qquad U=\exp(-ieQ\Theta)
\end{equation}
with $\Theta=\Theta(x)$ an arbitrary real function and the
electric charge matrix $Q={\rm diag}(2/3,-1/3,-1/3)$.
Writing out (\ref{phiu1}) for the components of $\Phi$, one finds that
each meson transforms with its proper electric charge as the generator.
With $A'_\mu=A_\mu-\partial_\mu\Theta$, the covariant derivative that
ensures electromagnetic gauge invariance is
\begin{equation}\label{dsigcov}
D_\mu\Sigma=\partial_\mu\Sigma-ieA_\mu\left[Q,\Sigma\right]
\end{equation}
Using this assignment, (\ref{lqcd2}) and (\ref{lds12}) include the
electromagnetic interactions of the mesons. At higher orders in the
chiral Lagrangian also terms with factors of the electromagnetic field
strengths $F_{\mu\nu}$ have to be included. In the chiral counting
$F_{\mu\nu}$ is equivalent to two powers of momentum.

We finally give a further illustration of the workings of $\chi$PT
with two examples of long-distance dominated, radiative kaon decays.
We first mention an important theorem for these processes.
It says that the amplitudes of nonleptonic radiative kaon decays with
{\it at most one pion\/} in the final state start only at ${\cal O}(p^4)$
in $\chi$PT.
This means there are no tree-level contributions at ${\cal O}(p^2)$.
Such terms are forbidden by gauge invariance.
There can only be tree-level amplitudes from ${\cal O}(p^4)$,
and, at the same order, loop contributions generated from ${\cal O}(p^2)$ 
interactions. Decays that fall under this category are
$K\to\gamma\gamma$, $K\to\gamma l^+l^-$, $K\to\pi\gamma\gamma$ or
$K\to\pi l^+l^-$.

\begin{figure}[t]
\hspace*{2cm}\epsfig{figure=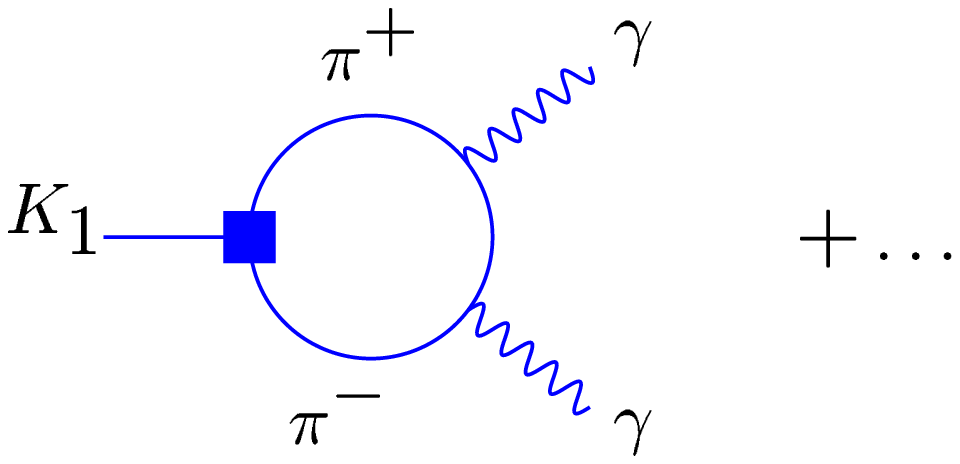,width=7cm,height=3cm}
\vspace*{-0cm}
\caption{$K_1\to\gamma\gamma$ in $\chi$PT. \label{fig:ksgg}}
\end{figure}

A particularly interesting example is $K_S\to\gamma\gamma$. In this case
it turns out that there is no direct coupling even at ${\cal O}(p^4)$
and hence no counterterm to absorb any divergence from the loop contribution.
As a consequence, the one-loop calculation (Fig. \ref{fig:ksgg})
is in fact
finite. The only parameter involved is $g_8$, which we have already
determined. The finite loop calculation then gives a unique 
prediction \cite{DEG}.
It yields
\begin{equation}\label{bksggth}
B(K_S\to\gamma\gamma)= 2.1\cdot 10^{-6}
\end{equation}
This compares well with the experimental result \cite{PDG}
$(2.4\pm 0.9)\cdot 10^{-6}$, which has recently been
improved to \cite{LAI} $(2.6\pm 0.4)\cdot 10^{-6}$.

\begin{figure}[h]
\hspace*{2cm}\epsfig{figure=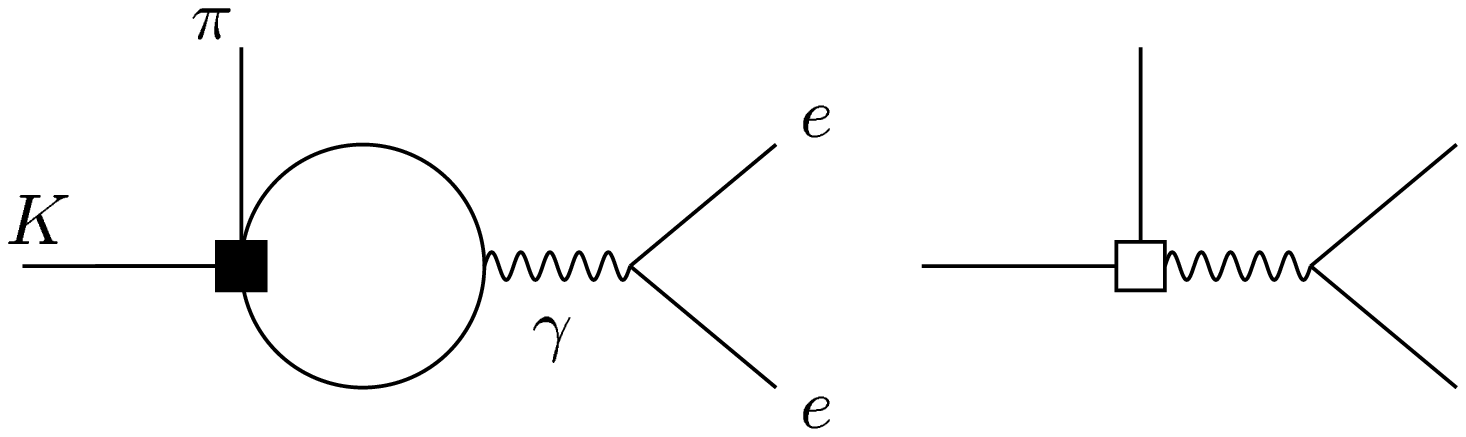,width=8cm,height=2.8cm}
\caption{$K^+\to\pi^+ e^+e^-$ in $\chi$PT. \label{fig:kpppee}}
\end{figure}

Of course this situation with a finite loop result is somewhat
special. A more generic case is $K^+\to\pi^+ e^+e^-$,
shown in Fig. \ref{fig:kpppee}.
Here the loop calculation is divergent and renormalized by the
counterterm at ${\cal O}(p^4)$. There is now one additional free
parameter, which can be determined from the rate of $K^+\to\pi^+ e^+e^-$.
Other observables as the $e^+e^-$ mass spectrum or the rate and spectrum
of $K^+\to\pi^+ \mu^+\mu^-$ can then be predicted.
In the same manner one can also analyze the
amplitude for $K_S \to \pi^0 e^+ e^-$, which determines the
indirect CP violating contribution in $K_L \to \pi^0 e^+ e^-$.
$K_S \to \pi^0 e^+ e^-$ is very similar to $K^+\to\pi^+ e^+e^-$, but
the required counterterm is different. For this reason the measurement
of $K^+\to\pi^+ e^+e^-$ cannot be used to obtain a prediction
for $K_S \to \pi^0 e^+ e^-$. 
A separate measurement of the latter decay will therefore be needed in
the future.


\section{The Neutral-$K$ System and CP Violation}

\subsection{Basic Formalism}

\begin{figure}[t]
\epsfig{figure=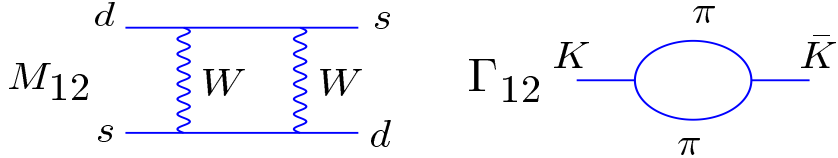,width=11.5cm,height=4.5cm}
\vspace*{-2cm}
\caption{Diagrams contributing to $M_{12}$ and $\Gamma_{12}$ in
the neutral kaon system.}
\label{fig:mg12}
\end{figure}

Neutral $K$ mesons can mix with their antiparticles through
second order weak interactions. They form a two-state system
($K^0-\bar K^0$) that is described
by a Hamiltonian matrix $\hat H$ of the form
\begin{equation}\label{gbuc:hmg}
\hat H= \left( \begin{array}{cc}
           M_{11} & M_{12}  \\  
           M^*_{12} & M_{11} 
         \end{array} \right) 
-\frac{i}{2} \left( \begin{array}{cc}
           \Gamma_{11} & \Gamma_{12}  \\  
           \Gamma^*_{12} & \Gamma_{11} 
         \end{array} \right) 
\end{equation}
where CPT invariance has been assumed. The absorptive part
$\Gamma_{ij}$ of $\hat H$ accounts for the weak decay of
the neutral kaon. In Fig. \ref{fig:mg12}
we show the diagrams that give rise to the off-diagonal elements
of $\hat H$.
Diagonalizing the Hamiltonian $\hat H$ yields the physical
eigenstates $K_{L,S}$. They are linear combinations of the strong
interaction eigenstates $K$ and $\bar K$ and can be written as
\begin{equation}\label{gbuc:fh}
K_L={\cal N}_{\bar\varepsilon}\left[(1+\bar\varepsilon)K +
  (1-\bar\varepsilon)\bar K\right]\equiv p K+ q \bar K
\end{equation}
\begin{equation}\label{gbuc:fl}
K_S={\cal N}_{\bar\varepsilon}\left[(1+\bar\varepsilon)K -
  (1-\bar\varepsilon)\bar K\right]\equiv p K - q \bar K
\end{equation}
with the normalization factor 
${\cal N}_{\bar\varepsilon}=1/\sqrt{2(1+|\bar\varepsilon|^2)}$.
Here $\bar\varepsilon$ is determined by
\begin{equation}\label{gbuc:epsmg}
\frac{1-\bar\varepsilon}{1+\bar\varepsilon}\equiv\frac{q}{p}=
\frac{M^*_{12}-\frac{i}{2}\Gamma^*_{12}}{\left(\Delta M
 +\frac{i}{2}\Delta\Gamma\right)/2}
\end{equation}
where $\Delta M$ and $\Delta\Gamma$ are the differences of the
eigenvalues $M_{L,S}-i\Gamma_{L,S}/2$ corresponding to the
eigenstates $K_{L,S}$
\begin{equation}\label{dmdg}
\Delta M\equiv M_L-M_S > 0 \qquad\qquad 
\Delta\Gamma\equiv \Gamma_S-\Gamma_L > 0
\end{equation}
The labels $L$ and $S$ denote, respectively, the long-lived and the 
short-lived eigenstate so that $\Delta\Gamma$ is positive by definition.
We employ the CP phase convention $CP\cdot K=-\bar K$.
Using the SM results for $M_{12}$, $\Gamma_{12}$ and standard
phase conventions  for the CKM matrix (see (\ref{gbuc:ckm})), one
finds in the limit of CP conservation ($\eta=0$) that
$\bar\varepsilon=0$. With (\ref{gbuc:fh}), (\ref{gbuc:fl}) it follows
that $K_L$ is CP odd and $K_S$ is CP even in this limit, which is close
to realistic since CP violation is a small effect.
As we shall see explicitly later on, the real part of $\bar\varepsilon$
is a physical observable, while the imaginary part is not. In particular
$(1-\bar\varepsilon)/(1+\bar\varepsilon)$ is a phase convention dependent,
unphysical quantity.

A crucial feature of the kaon system is the very large difference in 
decay rates between the two eigenstates, the lighter eigenstate
decaying much more rapidly than the heavier one
($\Gamma_S=579\,\Gamma_L$).
The basic reason is the small number of decay channels 
for the neutral kaons. Decay into the predominant
CP even two-pion final states $\pi^+\pi^-$, $\pi^0\pi^0$ is only
available for $K_S$, but not (to first approximation) for the
(almost) CP odd state $K_L$. The latter can decay into three pions,
which however is kinematically strongly suppressed, leading to a much
longer $K_L$ lifetime.

\subsection{Classification of CP Violation}

The fundamental weak interaction Lagrangian violates CP invariance
via the CKM mechanism, that is through an irreducible complex phase
in the quark mixing matrix. This
leads to a violation of CP symmetry at the phenomenological level, in
particular in the decays of mesons.
For instance, processes
forbidden by CP symmetry may occur or transitions related to each
other by CP conjugation may have a different rate.
In order for CP violation to manifest itself in this manner, an
interference of some sort between amplitudes is necessary. The
interference can arise in a variety of ways. It is therefore useful
to introduce a classification of the various possibilities.
We shall discuss it in terms of kaons, 
which are our main concern here, but it is applicable also
to $D$ and $B$ mesons in a similar way.
According to this classification, which is very common in the literature 
on CP violation, we may distinguish between:

a) {\it CP violation in the mixing matrix.\/}
This type of effect is based on CP violation in the two-state
mixing Hamiltonian $\hat H$ (\ref{gbuc:hmg}) itself and is measured by the
observable quantity ${\rm Im}(\Gamma_{12}/M_{12})$.
It is related to a change in flavour by two units,
$\Delta S = 2$.

b) {\it CP violation in the decay amplitude.\/} This class of phenomena
is characterized by CP violation originating directly in the amplitude
for a given decay. It is entirely independent of particle-antiparticle
mixing and can therefore occur for charged mesons 
as well. Here the transitions have $\Delta S =1$. 

c) {\it CP violation in the interference of mixing and decay.\/}
In this case the interference of two amplitudes
takes place between the mixing amplitude
and the decay amplitude in decays of neutral mesons.
This very important class is sometimes also refered to as
{\it mixing-induced\/} CP violation, a terminology not to be confused
with a).

Complementary to this classification is the widely used notion of
{\it direct\/} versus {\it indirect\/} CP violation. It is motivated
historically by the hypothesis of a new superweak interaction
which was proposed as early as 1964
by Wolfenstein to account for the CP violation observed in 
$K_L\to\pi^+\pi^-$ decay (see the lectures by Wolfenstein in this
volume). 
This new CP violating interaction would lead
to a local four-quark vertex that changes the flavour quantum number 
(strangeness) by two units. Its only effect would be a CP
violating contribution to $M_{12}$, so that all observed CP violation
could be attributed to particle-antiparticle mixing alone.
Today, after the advent of the three generation SM, the CKM mechanism
of CP violation appears more natural. In principle the superweak
scenario represents a logical possibility, leading to a different pattern
of observable CP violation effects. 
\\
Now, any CP violating effect that can be entirely assigned to CP
violation in $M_{12}$ (as for the superweak case) is termed
{\it indirect CP violation\/}. Conversely, any effect that can not be 
described in this way and explicitly requires CP violating phases in
the decay amplitude itself is called {\it direct CP violation\/}.
It follows that class a) represents indirect, class b) direct CP
violation. Class c) contains aspects of both. In this latter case
the magnitude of CP violation observed in any one decay mode (within the
neutral kaon system, say) could by itself be ascribed to mixing, thus
corresponding to an indirect effect. On the other hand, a difference in
the degree of CP violation between two different modes would reveal
a direct effect.

We illustrate these classes by a few important examples.
We will also use this opportunity to discuss several aspects
of kaon CP violation in more detail. 

\subsection*{a) -- Lepton Charge Asymmetry}

The lepton charge asymmetry in semileptonic $K_L$ decay is an 
example for CP violation in the mixing matrix. It is probably
the most obvious manifestation of CP nonconservation in kaon decays.
The observable considered here reads ($l=e$ or $\mu$)
\begin{eqnarray}\label{gbuc:lca}
\Delta &=& \frac{\Gamma(K_L\to\pi^-l^+\nu)-\Gamma(K_L\to\pi^+l^-\bar\nu)}{
  \Gamma(K_L\to\pi^-l^+\nu)+\Gamma(K_L\to\pi^+l^-\bar\nu)}=
\frac{|1+\bar\varepsilon|^2-|1-\bar\varepsilon|^2}{
   |1+\bar\varepsilon|^2+|1-\bar\varepsilon|^2} \nonumber \\
&\approx& 2{\rm Re}\ \bar\varepsilon
\approx\frac{1}{4}{\rm Im}\frac{\Gamma_{12}}{M_{12}}
\end{eqnarray}
If CP was a good symmetry of nature, $K_L$ would be a CP eigenstate and
the two processes compared in (\ref{gbuc:lca}) were related by a
CP transformation. The rate difference $\Delta$ should vanish.
Experimentally one finds however \cite{PDG}
\begin{equation}\label{gbuc:dexp}
\Delta_{exp}=(3.27\pm 0.12)\cdot 10^{-3}
\end{equation}
a clear signal of CP violation. The second equality in (\ref{gbuc:lca})
follows from (\ref{gbuc:fh}), noting that the
positive lepton $l^+$ can only originate from $K\sim(\bar sd)$,
$l^-$ only from $\bar K\sim (\bar ds)$. This is true to leading order
in SM weak interactions and holds to sufficient accuracy for our purpose.
The charge of the lepton essentially serves to tag the strangeness
of the $K$, thus picking out either only the $K$ or only the $\bar K$
component. Any phase in the semileptonic amplitudes is irrelevant
and the CP violation effect is purely in the mixing matrix itself.
In fact, as indicated in (\ref{gbuc:lca}), $\Delta$ is determined by
${\rm Im}(\Gamma_{12}/M_{12})$, the physical measure of CP
violation in the mixing matrix.
\\
{}From (\ref{gbuc:dexp}) we see that $\Delta>0$. This empirical fact
can be used to define positive electric charge in an absolute
sense. Positive charge is the charge of the lepton more copiously
produced in semileptonic $K_L$ decay. This definition is unambiguous
and would even hold in an antimatter world. Also, using 
some parity violation experiment, this result implies in addition an
absolute definition of left and right. These are quite remarkable
facts. They clearly provide part of the motivation to try to learn
more about the origin of CP violation.

\subsection*{b) -- CP Violation in the Decay Amplitude}

Observable CP violation may also occur through interference effects 
in the decay amplitudes themselves (pure direct CP violation).
This case is conceptually perhaps the simplest mechanism for
CP violation and the basic features are here particularly transparent.
Consider a situation where two different components contribute to the
amplitude of a $K$ meson decaying into a final state $f$
\begin{equation}\label{gbuc:akf}
A\equiv A(K\to f)=A_1 e^{i\delta_1}e^{i\phi_1}+
                  A_2 e^{i\delta_2}e^{i\phi_2}
\end{equation}
Here $A_i$ ($i=1,2$) are real amplitudes and $\delta_i$ are complex
phases from CP conserving interactions. The $\delta_i$ are usually
strong interaction rescattering phases. Finally the $\phi_i$ are weak
phases, that is phases coming from the CKM matrix in the SM.
The corresponding amplitude for the CP conjugated process
$\bar K\to\bar f$ then reads (the explicit minus signs are due to
our convention $CP\cdot K=-\bar K$, ($CP\cdot f=\bar f$))
\begin{equation}\label{gbuc:akfb}
\bar A\equiv A(\bar K\to \bar f)=-A_1 e^{i\delta_1}e^{-i\phi_1}-
                  A_2 e^{i\delta_2}e^{-i\phi_2}
\end{equation}
Since now all quarks are replaced by antiquarks (and vice versa) compared
to (\ref{gbuc:akf}), the weak phases change sign. The CP invariant
strong phases remain the same.
From (\ref{gbuc:akf}) and (\ref{gbuc:akfb}) one finds immediately
\begin{equation}\label{gbuc:aab}
|A|^2-|\bar A|^2 \sim A_1 A_2 \sin(\delta_1-\delta_2)
  \sin(\phi_1-\phi_2)
\end{equation}
The conditions for a nonvanishing difference between the decay rates
of $K\to f$ and the CP conjugate $\bar K\to\bar f$, that is direct
CP violation, can be read off from (\ref{gbuc:aab}).
There need to be two interfering amplitudes $A_1$, $A_2$ and these
amplitudes must simultaneously have both different weak ($\phi_i$) and
different strong phases ($\delta_i$). Although the strong interaction
phases can of course not generate CP violation by themselves, they are
still a necessary requirement for the weak phase differences to show up
as observable CP asymmetries. It is obvious from (\ref{gbuc:akf}) and
(\ref{gbuc:akfb}) that in the absence of strong phases $A$ and $\bar A$
would have the same absolute value despite their different weak phases,
since then $A=-\bar A^*$.

A specific example is given by the decays $K(\bar K)\to\pi^+\pi^-$
(here $f=\pi^+\pi^-=\bar f$). The amplitudes can be written as
\begin{eqnarray}\label{gbuc:apm}
A_{+-} &=& \sqrt{\frac{2}{3}}A_0 e^{i\delta_0}+
  \frac{1}{\sqrt{3}}A_2 e^{i\delta_2} \nonumber \\
\bar A_{+-} &=& -\sqrt{\frac{2}{3}}A^*_0 e^{i\delta_0}-
  \frac{1}{\sqrt{3}}A^*_2 e^{i\delta_2} 
\end{eqnarray}
where $A_{0,2}$
are the transition amplitudes of $K$ to the isospin-0 and
isospin-2 components of the $\pi^+\pi^-$ final state, defined by
\begin{equation}\label{a02def}
\langle\pi\pi(I=0,2)|{\cal H}_W|K\rangle\equiv
A_{0,2}e^{i\delta_{0,2}}
\end{equation}
They still
include the weak phases, but the strong phases have been factored out and 
written explicitly in (\ref{gbuc:apm}), (\ref{a02def}). 
Taking the modulus squared of the
amplitudes we get
\begin{eqnarray}\label{gbuc:reep}
\frac{\Gamma(K\to\pi^+\pi^-)-\Gamma(\bar K\to\pi^+\pi^-)}{
  \Gamma(K\to\pi^+\pi^-)+\Gamma(\bar K\to\pi^+\pi^-)} &=&
\sqrt{2}\sin(\delta_0-\delta_2)\frac{{\rm Re}A_2}{{\rm Re}A_0}
\left(\frac{{\rm Im}A_2}{{\rm Re}A_2}-\frac{{\rm Im}A_0}{{\rm Re}A_0}
\right) \nonumber \\
&=& 2\ {\rm Re}\ \varepsilon'
\end{eqnarray}
The quantity so defined is just twice the real part of the famous 
parameter $\varepsilon'$, the measure of direct CP violation in
$K\to\pi\pi$ decays.
The real parts of $A_{0,2}$ can be extracted from experiment.
The imaginary parts have to be calculated using the effective
Hamiltonian formalism.

We should stress that the quantity in (\ref{gbuc:reep}) is not the
observable actually used to determine $\varepsilon'$ experimentally.
We have discussed it here because it is of conceptual interest as
the simplest manifestation of $\varepsilon'$. The realistic analysis
requires a more general consideration of $K_L, K_S\to\pi\pi$ decays
to which we turn in the following paragraph.

\subsection*{c) -- Mixing Induced CP Violation in $K\to\pi\pi$:
  $\varepsilon$, $\varepsilon'$}

In this section we will illustrate the concept of mixing-induced
CP violation with the example of $K\to\pi\pi$ decays. These are
important processes, since CP violation has first been seen in
$K_L\to\pi^+\pi^-$ and as of today our most precise experimental
knowledge about this phenomenon still comes from the study of
$K\to\pi\pi$ transitions. There are two distinct final states and in a
strong interaction eigenbasis the transitions are 
$K^0,\bar K^0\to\pi\pi(I=0), \pi\pi(I=2)$, with definite isospin for
$\pi\pi$. Alternatively, using the physical eigenbasis for both
initial and final states, one has $K_L,K_S\to\pi^+\pi^-,\pi^0\pi^0$.

Consider next the amplitude for $K_L$ going into the CP even state
$\pi\pi(I=0)$, which can proceed via $K$ 
($\sim(1+\bar\varepsilon)A_0$) or via $\bar K$
($\sim(1-\bar\varepsilon)A^*_0$). Hence (to first order in small
quantities)
\begin{equation}\label{gbuc:ebe}
A(K_L\to\pi\pi(I=0))\sim 
(1+\bar\varepsilon)A_0 e^{i\delta_0}-
(1-\bar\varepsilon)A^*_0 e^{i\delta_0}\sim
\bar\varepsilon+i\frac{{\rm Im}A_0}{{\rm Re}A_0}=\varepsilon
\end{equation}
This defines the parameter $\varepsilon$, characterizing mixing-induced
CP violation. Note that $\varepsilon$ involves a component from
mixing ($\bar\varepsilon$) as well as from the decay amplitude
(${\rm Im}A_0/{\rm Re}A_0$). Neither of those is physical separately,
but $\varepsilon$ is. Note also that the physical quantity
${\rm Re}\,\bar\varepsilon$ discussed above satisfies
${\rm Re}\,\bar\varepsilon={\rm Re}\,\varepsilon$.
More generally one can form the following two CP violating observables
\begin{equation}\label{gbuc:epm0}
\eta_{+-}=\frac{A(K_L\to\pi^+\pi^-)}{A(K_S\to\pi^+\pi^-)}
\qquad\qquad
\eta_{00}=\frac{A(K_L\to\pi^0\pi^0)}{A(K_S\to\pi^0\pi^0)}
\end{equation}
These amplitude ratios involve the physical initial and final states
and are directly measurable in experiment. They are related to
$\varepsilon$ and $\varepsilon'$ through
\begin{equation}\label{gbuc:eteps}
\eta_{+-}=\varepsilon+\varepsilon'\qquad\qquad
\eta_{00}=\varepsilon -2\varepsilon'
\end{equation}
The phase of $\varepsilon$ is given by
$\varepsilon=|\varepsilon|\exp(i\pi/4)$. The relative phase between
$\varepsilon'$ and $\varepsilon$ can be determined theoretically.
It is close to zero so that to very good approximation
$\varepsilon'/\varepsilon={\rm Re}(\varepsilon'/\varepsilon)$.

Both $\eta_{+-}$ and $\eta_{00}$ measure mixing-induced CP violation
(interference between mixing and decay). Each of them considered 
separately could be attributed to CP violation in $K$--$\bar K$ mixing
and would therefore represent indirect CP violation. On the other hand,
a nonvanishing difference $\eta_{+-}-\eta_{00}=3\varepsilon'\not=0$
is a signal of direct CP violation.
Experimentally one has \cite{PDG}
\begin{equation}\label{gbuc:eps}
|\varepsilon|=(2.282\pm 0.019)\cdot 10^{-3}
\end{equation}

The quantity $\varepsilon'$ can be measured as the ratio
${\rm Re}(\varepsilon'/\varepsilon)\approx\varepsilon'/\varepsilon$
using the double ratio of rates
\begin{equation}\label{repe}
\left|\frac{\eta_{+-}}{\eta_{00}}\right|^2\doteq
1+6\ {\rm Re}\frac{\varepsilon'}{\varepsilon}
\end{equation}
Ten years ago, and until recently,
the experimental situation was characterized by the following,
somewhat inconclusive results \cite{BUR,GIB}: 
\begin{equation}\label{gbuc:epex1}
{\rm Re}\frac{\varepsilon'}{\varepsilon}=
\left\{ \begin{array}{ll}
        (23\pm 7)\cdot 10^{-4} & \mbox{CERN NA31} \\
        (7.4\pm 5.9)\cdot 10^{-4} & \mbox{FNAL E731}
        \end{array}
\right.
\end{equation}
In particular the second measurement was well compatible with
zero. A new round of experiments, conducted at both CERN and Fermilab,
was therefore anticipated with great interest.
The recent results have firmly established direct CP violation
\cite{ALA,FAN}:
\begin{equation}\label{gbuc:epex2}
{\rm Re}\frac{\varepsilon'}{\varepsilon}=
\left\{ \begin{array}{ll}
        (28.0\pm 4.1)\cdot 10^{-4} & \mbox{FNAL KTeV} \\
        (14.0\pm 4.3)\cdot 10^{-4} & \mbox{CERN NA48} 
        \end{array}
\right.
\end{equation}
These results rule out the superweak hypothesis, at least in its
most stringent form.
The analyses of the experiments are currently still ongoing
and should eventually settle the value of $\varepsilon'/\varepsilon$
to an accuracy of $(1-2)\cdot 10^{-4}$.

\subsection{Theory of $\varepsilon$ and the Unitarity Triangle} 

\subsubsection{Calculation of $\varepsilon$}

In the theoretical expression for $\varepsilon$ in (\ref{gbuc:ebe}),
the term ${\rm Im}A_0/{\rm Re}A_0$ is numerically negligible
(in standard phase convention). The value for $\varepsilon$ is
then approximately given by $\bar\varepsilon$ from (\ref{gbuc:epsmg}),
and can be written as
\begin{equation}\label{epsm12}
\varepsilon= e^{i\pi/4}\frac{{\rm Im}M_{12}}{\sqrt{2}\Delta M}
\end{equation}
$M_{12}$ is related to the first diagram shown in Fig. \ref{fig:mg12}. 
It is given by
\begin{equation}\label{m12ds2}
M_{12}=\frac{1}{2 m_K}
\langle K^0|{\cal H}^{\Delta S=2}_{eff}|\bar K^0\rangle
\end{equation}
Here ${\cal H}^{\Delta S=2}_{eff}$ is the effective Hamiltonian
for $\Delta S=2$ transitions, which is derived from the box
diagrams for $M_{12}$ in Fig. \ref{fig:mg12} by performing
an operator product expansion.
In this case there is only a single operator 
\begin{equation}\label{qds2}
Q^{\Delta S=2}=(\bar ds)_{V-A}(\bar ds)_{V-A}
\end{equation}
in the effective Hamiltonian.
One obtains explicitly
\begin{eqnarray}\label{epsth}
\varepsilon &=& e^{i\pi/4}\frac{G^2_F M^2_W f^2_K}{12\pi^2}
\frac{m_K}{\sqrt{2}\Delta M_K}B_K \nonumber\\
&& \cdot\mbox{Im}
\left[\lambda^{*2}_c S_0(x_c)\eta_1+\lambda^{*2}_t S_0(x_t)\eta_2
  +2\lambda^*_c\lambda^*_t S_0(x_c,x_t)\eta_3\right]
\end{eqnarray}
Here $\lambda_i=V^*_{is}V_{id}$, $f_K=160\,MeV$ is the kaon decay
constant and, at NLO, the bag parameter $B_K$ is defined by
\begin{equation}\label{bkrsi}
B_K=B_K(\mu)[\alpha^{(3)}_s(\mu)]^{-2/9}
 \left[1+\frac{\alpha^{(3)}_s(\mu)}{4\pi}J_3\right]
\end{equation}
\begin{equation}\label{bkme}
\langle K^0|(\bar ds)_{V-A}(\bar ds)_{V-A}|\bar K^0\rangle
  \equiv\frac{8}{3}B_K(\mu) f^2_K m^2_K
\end{equation}
The index $(3)$ in eq. (\ref{bkrsi}) refers to the number of flavours
in the effective theory and $J_3=307/162$ 
(in the so-called NDR scheme\cite{BBL}).
\\
The Wilson coefficient multiplying $B_K$ in (\ref{epsth}) consists
of a charm contribution, a top contribution and a mixed top-charm
contribution. It depends on the quark masses, $x_i\equiv m^2_i/M^2_W$,
through the functions $S_0$. The $\eta_i$ are the corresponding
short-distance QCD correction factors (which depend only slightly on
quark masses). Detailed definitions can be found in \cite{BBL}.
Numerical values for $\eta_1$, $\eta_2$ and $\eta_3$ are summarized
in Table \ref{etaitab}.
\begin{table}
\centering
\caption{NLO results for $\eta_i$ with
$\Lambda^{(4)}_{\overline{MS}}=(325\pm 110)\,MeV$,
$m_c(m_c)=(1.3\pm 0.05)\,GeV$, $m_t(m_t)=(170\pm 15)\,GeV$.
The third column shows the uncertainty due to the errors in
$\Lambda_{\overline{MS}}$ and quark masses. The fourth column
indicates the residual renormalization scale uncertainty at NLO
in the product of $\eta_i$ with the corresponding mass dependent function
from eq. (\ref{epsth}). These products are scale independent up to the
order considered in perturbation theory. The central values of the
QCD factors at LO are also given for comparison.}
\vskip 0.1 in
\begin{tabular}{|c|c|c|c|c|c|} \hline
& NLO(central) & $\Lambda_{\overline{MS}}$, $m_q$ &
scale dep. & NLO ref. & LO(central) \\
\hline
\hline
$\eta_1$ & 1.38 & $\pm 35\%$ & $\pm 15\%$ & \cite{HN1} & 1.12 \\
\hline
$\eta_2$ & 0.574 & $\pm 0.6\%$ & $\pm 0.4\%$ & \cite{BJW} & 0.61 \\
\hline
$\eta_3$ & 0.47 & $\pm 3\%$ & $\pm 7\%$ & \cite{HN3} & 0.35 \\
\hline
\end{tabular}
\label{etaitab}
\end{table}

Concerning these results the following remarks should be made.
\begin{itemize}
\item
$\varepsilon$ is dominated by the top contribution ($\sim 70\%$).
It is therefore rather satisfying that the related short distance
part $\eta_2 S_0(x_t)$ is theoretically extremely well under
control, as can be seen in Table \ref{etaitab}. Note in
particular the very small scale ambiguity at NLO, $\pm 0.4\%$
(for $100\,GeV\leq\mu_t\leq 300\,GeV$). This intrinsic theoretical
uncertainty is much reduced compared to the leading order result
where it would be as large as $\pm 9\%$.
\item
The $\eta_i$ factors and the hadronic matrix element are not
physical quantities by themselves. When quoting numbers it is therefore
essential that mutually consistent definitions are employed.
The factors $\eta_i$ described here are to be used in conjunction
with the so-called scale- (and scheme-) invariant bag parameter
$B_K$ introduced in (\ref{bkrsi}). The last factor on the 
right-hand side of
(\ref{bkrsi}) enters only at NLO. As a numerical example, if the
(scale and scheme dependent) parameter $B_K(\mu)$ is given in the
NDR scheme at $\mu=2GeV$, then (\ref{bkrsi}) becomes
$B_K=B_K(NDR,2\,GeV)\cdot 1.31\cdot 1.05$.
\item
The quantity $B_K$ has to be calculated by non-perturbative
methods. 
A representative range is
\begin{equation}
B_K=0.80\pm 0.15
\end{equation}
The status of $B_K$ is reviewed in \cite{BBGJJLS}.
\end{itemize}

\subsubsection{Determination of the Unitarity Triangle}

The source of CP violation in the standard model (SM) is the
Cabibbo-Kobayashi-Maskawa (CKM) matrix $V$ entering the charged-current
weak interaction Lagrangian
\begin{equation}\label{gbuc:lcc}
{\cal L}_{CC}=\frac{g_W}{2\sqrt{2}} V_{ij} \bar u_i\gamma^\mu
(1-\gamma_5) d_j W^+_\mu + h.c.
\end{equation}
where $(u_1, u_2, u_3)\equiv (u,c,t)$, 
$(d_1, d_2, d_3)\equiv (d,s,b)$ are the mass eigenstates of the
six quark flavours and a summation over $i,j=1,2,3$ is understood.

The unitary CKM matrix has the following explicit form
\begin{equation}\label{gbuc:ckm}
V= \left( \begin{array}{ccc}
           V_{ud} & V_{us} & V_{ub} \\ 
           V_{cd} & V_{cs} & V_{cb} \\ 
           V_{td} & V_{ts} & V_{tb} 
         \end{array} \right) 
\simeq
   \left( \begin{array}{ccc}
           1-\lambda^2/2 & \lambda & A\lambda^3(\varrho- i \eta) \\ 
           -\lambda & 1-\lambda^2/2 & A \lambda^2 \\ 
           A\lambda^3(1-\varrho-i \eta) & -A\lambda^2 & 1
         \end{array} \right) 
\end{equation} 
where the second expression is a convenient parametrization in
terms of $\lambda$, $A$, $\varrho$ and $\eta$ due to Wolfenstein.
It is organized as a series expansion in powers of $\lambda=0.22$
(the sine of the Cabibbo angle) to exhibit the hierarchy among
the transitions between generations. 
The explicit parametrization shown in (\ref{gbuc:ckm})
is valid through order ${\cal O}(\lambda^3)$, an approximation that is
sufficient for most practical applications. Higher order terms can
be taken into account if necessary \cite{BBL}.

\begin{figure}[t]
\hspace*{0.3cm}\epsfig{figure=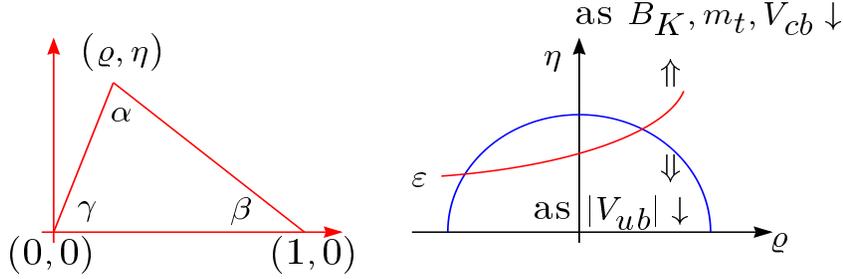,width=11.5cm,height=5.5cm}
\vspace*{-1.2cm}
\caption{The normalized unitarity triangle in the 
($\varrho$, $\eta$) plane (left), and
its standard determination (right).} 
\label{fig:uteps}
\end{figure}

The unitarity structure of the CKM matrix is conventionally displayed
in the so-called unitarity triangle, Fig. \ref{fig:uteps} (left).
This triangle is a graphical representation of the unitarity relation
$V_{ud}V^*_{ub}+V_{cd}V^*_{cb}+V_{td}V^*_{tb}=0$
(normalized by $-V_{cd}V^*_{cb}$) in the complex plane of Wolfenstein
parameters $(\varrho,\eta)$.
The angles $\alpha$, $\beta$ and $\gamma$ of the unitarity triangle
are phase convention independent and can be determined in CP violation
experiments. The area of the unitarity triangle, which is proportional
to $\eta$, is a measure of CP nonconservation in the standard model.

We briefly summarize the main ingredients of the standard analysis
of the unitarity triangle, where $\varepsilon$ plays a central role.
There are 4 independent parameters $\lambda$, $A$, $\varrho$ and $\eta$
in the CKM matrix, which are determined from 4 measurements as follows:
\begin{itemize}
\item
$\lambda=0.22$, from $K^+\to\pi^0 l^+\nu$, or from
semileptonic hyperon decay ($\Lambda\to pe\bar\nu$,   
$\Sigma^-\to ne\bar\nu$, $\Xi^-\to\Lambda e\bar\nu$).
\item
$V_{cb}=A\lambda^2=0.040\pm 0.002$, from $b\to cl\nu$ transitions.
\item
$|V_{ub}/V_{cb}|=\lambda\sqrt{\varrho^2+\eta^2}=0.09\pm 0.02$,
from $b\to ul\nu$ transitions.
\item
$|\varepsilon|\sim\eta\left((1-\varrho)A^2 S(m_t)+c\right)A^2 B_K$

with $B_K=0.80\pm 0.15$, from indirect CP violation in $K\to\pi\pi$.
\end{itemize}

Under the final item we have indicated the dependence of 
$\varepsilon$ (from (\ref{epsth})) on the most important parameters,
writing the CKM quantities explicitly in Wolfenstein form. 
Here $c$ denotes a constant that is independent of $A$, $\varrho$,
$\eta$, $m_t$ and $B_K$.

The standard determination of the unitarity triangle is
illustrated in Fig. \ref{fig:uteps} (right).
The relevant input parameters are
$B_K$, $m_t$, $V_{cb}$ and $|V_{ub}/V_{cb}|$. For fixed
$B_K$, $m_t$ and $V_{cb}$, the measured $|\varepsilon|$ determines
a hyperbola in the $\varrho$--$\eta$ plane of Wolfenstein 
parameters.
Intersecting the hyperbola with the circle defined by
$|V_{ub}/V_{cb}|$ determines the unitarity triangle (up to a two-fold
ambiguity). 
There is a simple regularity, which is quite useful and easy to
remember:
As {\it any\/} one of the four input parameters becomes {\it too small\/}
(with the others held fixed), the SM picture becomes inconsistent
(see Fig. \ref{fig:uteps}).
Using this fact, lower bounds on these parameters can be derived.
The large value that has been established for the
top-quark mass in fact helps to maintain the consistency of the SM.

In principle, once the unitarity triangle is fixed in this manner,
any further, independent measurement of a quantity in the
$(\varrho,\eta)$ plane provides us with an additional standard model
test. In practice, however, the accuracy of such a test is limited
by hadronic uncertainties, which enter mainly through $B_K$ and
$|V_{ub}/V_{cb}|$. 
Useful additional restrictions come from $B$--$\bar B$ mixing.
Both $\Delta M_d$ and $\Delta M_d/\Delta M_s$, where
$\Delta M_q$ is the mass difference in the $B_q$--$\bar B_q$
system, constrain $\sqrt{(1-\varrho)^2+\eta^2}$. 
The ratio $\Delta M_d/\Delta M_s$ is particularly important,
because the hadronic uncertainties cancel in the limit
of $SU(3)$ symmetry. Currently, while $\Delta M_d$ is well measured,
only a lower bound $\Delta M_s > 14.3\,{\rm ps}^{-1}$
exists at present. However, already this bound is interesting.
Together with $\Delta M_d$, it implies a quite clean upper bound
on $\sqrt{(1-\varrho)^2+\eta^2}$. This essentially excludes
negative values of $\varrho$, severely restricting the allowed
range of $\varrho$ and $\eta$.

\begin{figure}
\begin{center}
 \vspace{5.5cm}
\includegraphics{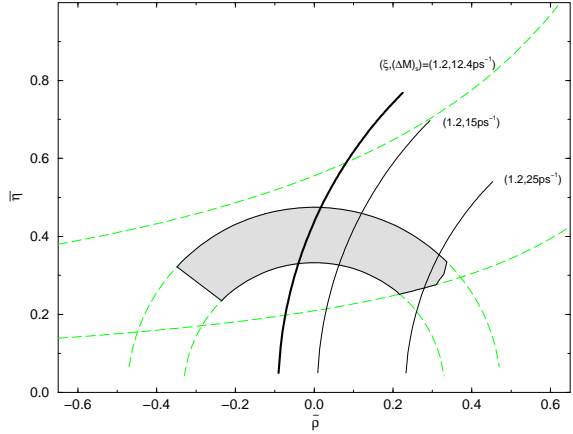}
\end{center} 
\caption{The allowed region (shaded) in the ($\bar\varrho$, $\bar\eta$) 
plane, combining information from $\varepsilon$, $|V_{ub}/V_{cb}|$
and including the constraint from $\Delta M_d$. The independent 
constraint from the lower limit on $\Delta M_s/\Delta M_d$
excludes the region to the left of the curves labeled with 
$\Delta M_s$ in the plot. $\xi\simeq 1.2$ measures $SU(3)$ breaking
in the hadronic matrix elements of 
$B_d$--$\bar B_d$ versus $B_s$--$\bar B_s$ mixing.  \label{fig:utnum}}
\end{figure}

The results of a complete analysis of the unitarity triangle
are shown in Fig. \ref{fig:utnum}. 
Here the axes are labeled
by $\bar\varrho=\varrho(1-\lambda^2/2)$ and 
$\bar\eta=\eta(1-\lambda^2/2)$, instead of $\varrho$ and $\eta$.
In this way higher terms in the Wolfenstein expansion $\sim\lambda^2$,
which can be neglected to first approximation,
are consistently taken into account.
The plot is taken from \cite{AJB99} where further details can be found.

\subsection{Calculating $\varepsilon'/\varepsilon$}

\begin{figure}[t]
\epsfig{figure=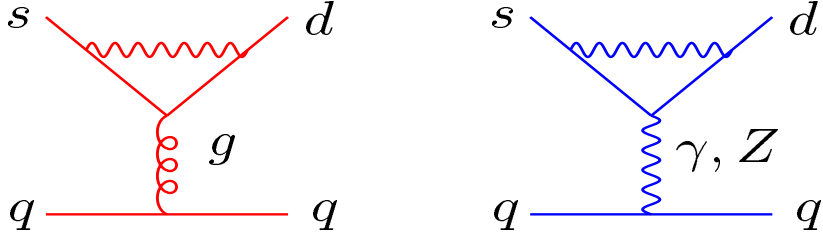,width=11.5cm,height=3.5cm}
\caption{Gluonic and electroweak penguin contributions,
which give rise to operators $Q_6$ and $Q_8$, respectively.}
\label{fig:epeq6q8}
\end{figure}

The formula for $\varepsilon'/\varepsilon$, which can be derived
from the definition in (\ref{gbuc:epm0}), (\ref{gbuc:eteps}),
is given by
\begin{equation}\label{epea02}
\frac{\varepsilon'}{\varepsilon}=\frac{\omega}{\sqrt{2}
  |\varepsilon|}\left(\frac{\mbox{Im} A_2}{\mbox{Re} A_2}-
        \frac{\mbox{Im} A_0}{\mbox{Re} A_0}\right)
\end{equation}
where $\omega\equiv\mbox{Re} A_2/\mbox{Re} A_0$.
This may be compared with (\ref{gbuc:reep}) using
\begin{equation}\label{argeps}
{\rm arg}(\varepsilon')=\frac{\pi}{2}+\delta_2-\delta_0
\approx \frac{\pi}{4}
\end{equation}
The expression (\ref{epea02}) for $\varepsilon'/\varepsilon$ may
also be written as
\begin{equation}\label{epea02b}
\frac{\varepsilon'}{\varepsilon}=-\frac{\omega}{\sqrt{2}
  |\varepsilon|\mbox{Re} A_0}\left(\mbox{Im} A_0-
  \frac{1}{\omega}\mbox{Im} A_2\right)
\end{equation}
$\mbox{Im} A_{0,2}$ are calculated from the general low energy
effective Hamiltonian for $\Delta S=1$ transitions (\ref{hds1}),
which we have described in sec. 3.4.
One has
\begin{equation}\label{ima02}
\mbox{Im} A_{0,2}=-\mbox{Im}\lambda_t \frac{G_F}{\sqrt{2}}
 \sum^{10}_{i=3} y_i(\mu)\langle Q_i\rangle_{0,2}
\end{equation}
Here $y_i$ are the Wilson coefficients and
$\langle Q_i\rangle_{0,2}e^{i\delta_{0,2}}
\equiv\langle\pi\pi(I=0,2)|Q_i|K^0\rangle$,
$\lambda_t=V^*_{ts}V_{td}$.
\\
For the purpose of illustration we keep only the numerically
dominant contributions and write
\begin{equation}\label{epeapr}
\frac{\varepsilon'}{\varepsilon}=
\frac{\omega G_F}{2|\varepsilon|\mbox{Re} A_0}\mbox{Im}\lambda_t
\left(y_6\langle Q_6\rangle_0-\frac{1}{\omega}y_8\langle Q_8\rangle_2
+\ldots\right)
\end{equation}
$Q_6$ originates from gluonic penguin diagrams and $Q_8$ from
electroweak contributions, as indicated schematically in 
Fig. \ref{fig:epeq6q8}. 
The matrix elements of $Q_6$ and
$Q_8$ can be parametrized by bag parameters $B_6$ and $B_8$ as
\begin{equation}\label{q6me}
\langle Q_6\rangle_0 =-4\sqrt{\frac{3}{2}}
\left[\frac{m_K}{m_s(\mu)+m_d(\mu)}\right]^2 m^2_K(f_K-f_\pi)\cdot B_6
\sim \left(\frac{m_K}{m_s}\right)^2 B_6
\end{equation}
\begin{equation}\label{q8me}
\langle Q_8\rangle_2\simeq\sqrt{3}
\left[\frac{m_K}{m_s(\mu)+m_d(\mu)}\right]^2 m^2_K f_\pi\cdot B_8
\sim \left(\frac{m_K}{m_s}\right)^2 B_8
\end{equation}
$B_6=B_8=1$ corresponds to the factorization assumption for the
matrix elements, which holds in the large $N_C$ limit of QCD.

The numerical importance of the contributions from $Q_6$ and $Q_8$
can be understood as follows.
$Q_{6,8}$ are particular because they are of the
$(V-A)\otimes (V+A)$ form, which results in a $(S+P)\otimes (S-P)$
structure upon Fierz transformation. Factorizing the matrix element
of such operators gives for example
\begin{equation}\label{spfact}
\langle\pi^+\pi^-|(\bar su)_{S+P}(\bar ud)_{S-P}|K\rangle
\ \to\  -\langle\pi^-|\bar su|K\rangle
\cdot\langle\pi^+|\bar u\gamma_5 d|0\rangle
\end{equation}
Taking the derivative ($\partial^\mu$) of
\begin{equation}\label{udfpi}
\langle\pi^+(p)|(\bar u\gamma_\mu\gamma_5 d)(x)|0\rangle
=f_\pi p_\mu e^{ip\cdot x}
\end{equation}
and using the equations of motion, we find
\begin{equation}\label{u5d}
\langle\pi^+|\bar u\gamma_5 d|0\rangle=
f_\pi\frac{m^2_\pi}{m_u+m_d}=f_\pi\frac{m^2_K}{m_s+m_d}
\end{equation}
Here the second equality follows from the $\chi$PT relations
in (\ref{m2bmq}). A second factor of $m^2_K/(m_s+m_d)$ comes
in a similar way from the scalar current matrix element
$\langle\pi^-|\bar su|K\rangle$. This explains the
quark mass dependence in (\ref{q6me}) and (\ref{q8me}).
Since
\begin{equation}\label{mksdb0}
\frac{m^2_K}{m_s+m_d}=B_0={\cal O}(\Lambda_{QCD})
\end{equation}
we see that the matrix elements are primarily not proportional
to $(m_s+m_d)^{-2}$, but to $B_0$, which remains finite in the
chiral limit $m_s$, $m_d\to 0$. However $m_K$ is precisely known
and it is customary to trade the $\chi$PT parameter $B_0$ for the
quark masses $m_s+m_d\approx m_s$. Because $B_0$ is numerically,
and somewhat accidentally, quite large (equivalently, the quark masses
quite small), the matrix elements of $Q_{6,8}$ are systematically
enhanced over those of the ordinary $(V-A)\otimes (V-A)$ operators.
$Q_5$ and $Q_7$ have a Dirac structure similar to $Q_6$ and $Q_8$,
but a different colour structure, which leads to a $1/N_C$
suppression. In addition their Wilson coefficients are numerically
smaller. This implies that $Q_6$ and $Q_8$ give the dominant
contributions.

$y_6\langle Q_6\rangle_0$ and $y_8\langle Q_8\rangle_2$
are positive numbers. The value for $\varepsilon'/\varepsilon$
in (\ref{epeapr}) is thus characterized by a potential cancellation of
two competing contributions. Since the second contribution is an
electroweak effect, suppressed by $\sim\alpha/\alpha_s$ compared
to the leading gluonic penguin $\sim\langle Q_6\rangle_0$,
it could appear at first sight that it should be altogether
negligible for $\varepsilon'/\varepsilon$. However, a number of
circumstances actually conspire to systematically enhance the
electroweak effect so as to render it a sizable contribution:
\begin{itemize}
\item
Unlike $Q_6$, which is a pure $\Delta I=1/2$ operator,
$Q_8$ can give rise to the $\pi\pi(I=2)$ final state and thus
yield a non-vanishing $\mbox{Im} A_2$ in the first place.
\item
The ${\cal O}(\alpha/\alpha_s)$ suppression is largely compensated
by the factor $1/\omega\approx 22$ in (\ref{epeapr}), reflecting the
$\Delta I=1/2$ rule.
\item
$-y_8\langle Q_8\rangle_2$ gives a negative contribution to
$\varepsilon'/\varepsilon$ that strongly grows with $m_t$.
For the realistic top mass value it can be
substantial.
\end{itemize}

In order to estimate $\varepsilon'/\varepsilon$ numerically
(see (\ref{epeapr})), the hadronic matrix elements have to be 
determined within a nonperturbative framework
(e.g. lattice QCD, $1/N_C$ expansion, chiral quark model),
while the coefficients $y_i$ are known from perturbation theory,
and ${\rm Re}A_0$, $\omega$, $G_F$, $|\varepsilon|$ are fixed
from experiment. Finally, the CKM quantity 
${\rm Im}\lambda_t\sim\eta$
is obtained from the standard determination of the unitarity
triangle described in the previous section. 

The Wilson coefficients $y_i$ have been calculated at NLO
\cite{BJLW,CFMR}. The short-distance part is therefore quite
well under control. The remaining problem is then the computation
of matrix elements, in particular $B_6$ and $B_8$. The cancellation
between these contributions enhances the relative sensitivity of
$\varepsilon'/\varepsilon$ to the anyhow uncertain hadronic parameters
which makes a precise calculation of $\varepsilon'/\varepsilon$
impossible at present. 

The order of magnitude of $\varepsilon'$ can however be understood
from (\ref{epea02}). The size of ${\rm Im}A_i/{\rm Re}A_i$ is
essentially determined by the small CKM parameters that carry the complex 
phase and which are related to the top quark in the loop diagrams of
Fig. \ref{fig:epeq6q8}. Roughly speaking
${\rm Im}A_i/{\rm Re}A_i\sim {\rm Im}V^*_{ts}V_{td}\sim 10^{-4}$.
Empirically we have, from the $\Delta I=1/2$ rule,
${\rm Re}A_2/{\rm Re}A_0\sim 10^{-2}$. This leads to a natural size
of $\varepsilon'$ of $\sim 10^{-6}$, 
thus $\varepsilon'/\varepsilon\sim 10^{-3}$.

A complete analysis gives the result \cite{BGGJS,BBGJJLS}
\begin{equation}
1.4\cdot 10^{-4}\leq\varepsilon'/\varepsilon\leq 32.7\cdot 10^{-4} 
\ \ \ \mbox{(scanning)}
\end{equation}
\begin{equation}
5.2\cdot 10^{-4}\leq\varepsilon'/\varepsilon\leq 16.0\cdot 10^{-4} 
\ \ \ \mbox{(Gaussian)}
\end{equation}
This is compatible with the experimental results 
(\ref{gbuc:epex1}), (\ref{gbuc:epex2})
within the rather large uncertainties. 
The two ranges refer to different treatments of uncertainties
in the experimental input: the assumption of Gaussian errors, or 
flat distributions (scanning).
Similar findings are reported by other groups \cite{CM}$^{-}$\cite{WU}.
A detailed review of the theoretical status of $\varepsilon'/\varepsilon$,
including recent developments (hadronic matrix elements,
final state interactions, isospin breaking corrections etc.),
along with further references can be found in \cite{AJB01}.

The recent experimental confirmation that indeed $\varepsilon'\not= 0$
constitutes a qualitatively new feature of CP violation and
is as such of great importance. 
However, due to the large uncertainties in the
theoretical calculation, a quantitative use of this result for the
extraction of CKM parameters is unfortunately rather limited.
For this purpose one has to turn to theoretically cleaner
observables. As we will see in the next section, rare
$K$ decays in fact offer very promising opportunities in this direction.

\section{Rare $K$ Decays}

\subsection{$K^+\to\pi^+\nu\bar\nu$ and $K_L\to\pi^0\nu\bar\nu$}

\begin{figure}[t]
\epsfig{figure=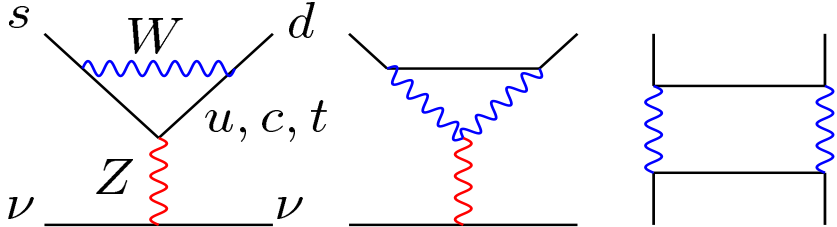,width=11.5cm,height=3.5cm}
\caption{The leading order electroweak diagrams contributing to
$K\to\pi\nu\bar\nu$ in the standard model.}
\label{fig:kpnn}
\end{figure}

The decays $K\to\pi\nu\bar\nu$ proceed through flavour changing
neutral currents. These arise in the standard model only at
second (one-loop) order in the electroweak interaction
(Z-penguin and W-box diagrams, Fig. \ref{fig:kpnn}) 
and are additionally GIM suppressed.
The branching fractions are thus very small, at the level of
$10^{-10}$, which makes a detection of these modes rather challenging.
However, the loop process $K\to\pi\nu\bar\nu$, a genuine quantum
effect of standard model flavour dynamics, probes important
short distance physics, in particular properties of the top quark
($m_t$, $V_{td}$, $V_{ts}$). It is also very sensitive to
potential new physics effects.
At the same time, the $K\to\pi\nu\bar\nu$ modes are reliably
calculable, in contrast to most other decay modes of interest.
A measurement of $K^+\to\pi^+\nu\bar\nu$ and $K_L\to\pi^0\nu\bar\nu$
will therefore be an extremely useful test of flavour physics.

Let us discuss the main properties of these decays, concentrating first
on the charged mode.
The GIM structure of the amplitude can be written as
\begin{equation}\label{gimf}
\sum_{i=u,c,t}\lambda_i\, F(x_i)=
\lambda_c\, (F(x_c)-F(x_u))+\lambda_t\, (F(x_t)-F(x_u))
\end{equation}
with $\lambda_i=V^*_{is}V_{id}$ and $x_i=m^2_i/M^2_W$.
The first important point is the characteristic {\it hard\/}
GIM cancellation pattern, which means that the function $F$
depends as a power on the internal mass scale
\begin{equation}\label{fuct}
F(x_u)\sim\frac{\Lambda^2_{QCD}}{M^2_W}\sim 10^{-5}
\ll F(x_c)\sim \frac{m^2_c}{M^2_W}\ln\frac{M_W}{m_c}\sim 10^{-3}
\ll F(x_t)\sim 1
\end{equation}
The up-quark contribution is a long-distance effect,
determined by the scale $\Lambda_{QCD}$.
As an immediate consequence, top and charm contribution
with their hard scales $m_t$, $m_c$
dominate the amplitude, whereas the long-distance part $F(x_u)$
is negligible. Note that the charm contribution,
$\lambda_c\, F(x_c)\sim 10^{-1}\cdot 10^{-3}$, and the top
contribution,  $\lambda_t\, F(x_t)\sim 10^{-4}\cdot 1$, have the
same order of magnitude when the CKM factors are included.

\begin{figure}[t]
\hspace*{0.4cm}\epsfig{figure=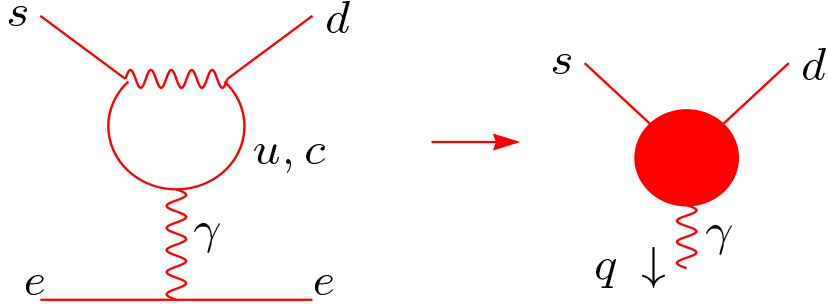,width=11.5cm,height=4.5cm}
\caption{Soft GIM mechanism in the photon penguin.}
\label{fig:gimsoft}
\end{figure}

The origin of the hard GIM mechanism is the fact that the neutrinos
only couple to the heavy gauge boson $W$ and $Z$. It is interesting
to contrast the situation with $K^+\to\pi^+ e^+e^-$, where photon
exchange can contribute as shown in Fig. \ref{fig:gimsoft}.
For simplicity we consider the case of internal quarks that are
light compared to $M_W$. The $W$ propagator can then be contracted
and the loop reduces essentially to a vacuum polarization diagram.
Electromagnetic gauge invariance, which is unbroken, requires the
$\bar sd$-photon vertex to have the form
\begin{equation}\label{gnuq}
\Gamma_\nu(q)\sim\bar s\gamma^\mu(1-\gamma_5)d\, 
\cdot (q^2 g_{\mu\nu}-q_\mu q_\nu)\, \ln\frac{M_W}{m_i}
\end{equation}
where $q$ is the photon momentum and we assumed
$q^2\ll m^2_i\ll M^2_W$. The structure in (\ref{gnuq}) ensures 
current conservation, $q^\nu\Gamma_\nu(q)\equiv 0$.
When the vertex is contracted with the electron current
$\bar e\gamma^\nu e$, the $q_\mu q_\nu$ term vanishes by the equations
of motion. The $q^2$ factor of the remaining term is canceled by the
photon propagator, which yields a local
$(\bar sd)_{V-A}(\bar ee)_V$ interaction and a loop function
\begin{equation}\label{fxiln}
F(x_i)\sim \ln\frac{M_W}{m_i}
\end{equation}
The logarithmic behaviour of the photon penguin is refered
to as the {\it soft\/} GIM mechanism. It is in contrast to the
hard GIM structure arising from the $W$ and $Z$ contributions
where gauge symmetry is spontaneously broken, leading to
the power behaviour $\sim (m^2_i/M^2_W)\ln(M_W/m_i)$.
The unsuppressed sensitivity to the light quark mass
$m_i=m_u$ in (\ref{fxiln})
is a signal of the long-distance dominance in the $K^+\to\pi^+ e^+e^-$
amplitude. Of course, $\chi$PT is needed for a consistent analysis
in this case. The quark-level result in (\ref{fxiln}), derived
in perturbation theory, is strictly speaking not valid, but it is
enough to indicate the long-distance sensitivity and the order of
magnitude of the contribution.

The short-distance dominance of the $s\to d\nu\bar\nu$ transition
next implies that the process is effectively semileptonic, because
a single, local operator $(\bar sd)_{V-A}(\bar\nu\nu)_{V-A}$
describes the interaction at low-energy scales.
Hence the amplitude has the form
\begin{equation}\label{akpnn}
A(K^+\to\pi^+\nu\bar\nu)\sim
G_F\alpha(\lambda_c F_c+\lambda_t F_t)
\langle\pi^+|(\bar sd)_V|K^+\rangle\, (\bar\nu\nu)_{V-A}
\end{equation}
The coefficient function $\lambda_c F_c+\lambda_t F_t$ is
calculable in perturbation theory. The hadronic matrix element
can be extracted from $K^+\to\pi^0 e^+\nu$ decay via the
isospin relation (\ref{kpiso}).
The $K^+\to\pi^+\nu\bar\nu$ amplitude is then completely determined,
and with good accuracy.

The neutral mode proceeds through CP violation in the standard model.
This is due to the definite CP properties of $K^0$, $\pi^0$ and
the hadronic transition current $(\bar sd)_{V-A}$. 
Using
\begin{equation}\label{cpkpi}
CP|\pi^0\rangle=-|\pi^0\rangle\qquad
CP|K^0\rangle=-|\bar K^0\rangle\qquad
CP\, (\bar sd)_V\, (CP)^{-1}=-(\bar ds)_V
\end{equation}
we have
\begin{equation}\label{pisdk}
\langle\pi^0|(\bar sd)_V|K^0\rangle=
-\langle\pi^0|(\bar ds)_V|\bar K^0\rangle
\end{equation}
(The axial vector currents $(\bar sd)_A$, $(\bar ds)_A$
do not contribute to the $K\to\pi$ transition because of parity.)
With $K_L=(K^0+\bar K^0)/\sqrt{2}$ ( the 
$\bar\varepsilon$-contribution is negligible) we then obtain for
the matrix element of the hadronic transition current
\begin{equation}\label{pisddskl}
\langle\pi^0|\lambda_i(\bar sd)_V+\lambda^*_i(\bar ds)_V|K_L\rangle
\sim {\rm Im}\,\lambda_i
\end{equation}
where $\lambda_i$ is the appropriate CKM factor.
This demonstrates the CP violating character of the leading
standard model amplitude for $K_L\to\pi^0\nu\bar\nu$.
For simplicity we have given the argument here assuming standard
phase conventions. A manifestly phase convention independent
derivation of the same result is discussed in \cite{BB6}.
The amplitude then has the form
\begin{equation}\label{aklpn}
A(K_L\to\pi^0\nu\bar\nu)\sim 
{\rm Im}\lambda_t\, F_t + {\rm Im}\lambda_c\, F_c
\end{equation}
where
\begin{equation}\label{imftc}
{\rm Im}\lambda_t\, F_t \sim 10^{-4}\cdot 1
\gg {\rm Im}\lambda_c\, F_c \sim 10^{-4}\cdot 10^{-3}
\end{equation}
The violation of
CP symmetry in $K_L\to\pi^0\nu\bar\nu$ arises through interference
between $K^0$--$\bar K^0$ mixing and the decay amplitude. This mechanism
is an example of mixing-induced CP violation. In the
standard model, the mixing-induced CP violation in $K_L\to\pi^0\nu\bar\nu$
is larger by orders of magnitude than the one in $K_L\to\pi^+\pi^-$,
for instance. 
This is because 
\begin{equation}
\frac{A(K_L\to\pi^0\nu\bar\nu)}{A(K_S\to\pi^0\nu\bar\nu)}={\cal O}(1)
\end{equation}
in contrast to the per-mille-size ratios in (\ref{gbuc:epm0}).
Any difference in the magnitude of mixing induced
CP violation between two $K_L$ decay modes is a signal of direct
CP violation. For this reason, the standard model decay 
$K_L\to\pi^0\nu\bar\nu$ is
a signal of almost pure direct CP violation, revealing an effect
that can not be explained by CP violation in the $K-\bar K$
mass matrix alone.

The $K\to\pi\nu\bar\nu$ modes have been studied in great detail
over the years to quantify the degree of theoretical precision.
Important effects come from short-distance QCD corrections.
These were computed at leading order in \cite{DDG}.
The complete next-to-leading order  calculations \cite{BB123,MU,BB99}
reduce the theoretical uncertainty in these decays to
$\sim 5\%$ for $K^+\to\pi^+\nu\bar\nu$ and $\sim 1\%$ for
$K_L\to\pi^0\nu\bar\nu$.
This picture is essentially unchanged when further small effects
are considered, including isospin breaking in the relation of
$K\to\pi\nu\bar\nu$ to $K^+\to\pi^0l^+\nu$ \cite{MP},
long-distance contributions
\cite{RS,HLLW}, the CP-conserving effect in $K_L\to\pi^0\nu\bar\nu$
in the standard model \cite{RS,BI}, two-loop electroweak 
corrections for large $m_t$ \cite{BB7} and subleading-power
corrections in the OPE in the charm sector \cite{FLP}.

While already $K^+\to\pi^+\nu\bar\nu$ can be reliably calculated,
the situation is even better for $K_L\to\pi^0\nu\bar\nu$. Since
only the imaginary part of the amplitude 
contributes, the charm sector, in $K^+\to\pi^+\nu\bar\nu$
the dominant source of uncertainty, is completely negligible for
$K_L\to\pi^0\nu\bar\nu$ ($0.1\%$ effect on the branching ratio).
Long distance contributions 
($\;\raisebox{-.4ex}{\rlap{$\sim$}} \raisebox{.4ex}{$<$}\; 0.1\%$)  
and also the indirect CP violation effect  
($\;\raisebox{-.4ex}{\rlap{$\sim$}} \raisebox{.4ex}{$<$}\; 1\%$)  
are likewise negligible. 
The total theoretical
uncertainties, from perturbation theory in the top sector
and in the isospin breaking corrections, are safely below
$3\%$ for $B(K_L\to\pi^0\nu\bar\nu)$. This makes this decay
mode truly unique and very promising for phenomenological
applications.

In Table \ref{kpnntab} we have summarized some of the main
features of $K^+\to\pi^+\nu\bar\nu$ and $K_L\to\pi^0\nu\bar\nu$.
\begin{table}
\centering
\caption{\it Compilation of important properties and results
for $K\to\pi\nu\bar\nu$.
}
\vskip 0.1 in
\begin{tabular}{|c|c|c|} \hline
& $K^+\to\pi^+\nu\bar\nu$ & $K_L\to\pi^0\nu\bar\nu$ \\
\hline
\hline
& CP conserving & CP violating \\
\hline
CKM & $V_{td}$ & $\mbox{Im} V^*_{ts}V_{td}\sim J_{CP}\sim\eta$ \\
\hline
contributions & top and charm & only top \\
\hline
scale dep. (BR) &  $\pm 20\%$ (LO) &
                   $\pm 10\%$ (LO) \\
                &  $\to\pm 5\%$ (NLO) &
                   $\to\pm 1\%$ (NLO) \\
\hline
BR (SM) & $(0.8\pm 0.3)\cdot 10^{-10}$&$(2.8\pm 1.1)\cdot 10^{-11}$ \\
\hline
exp.  & $\left(1.5^{+3.4}_{-1.2}\right)\cdot 10^{-10}$ BNL 787 \cite{SAT}
           & $< 5.9\cdot 10^{-7}$ KTeV \cite{KLPNTEV} \\
\hline
\end{tabular}
\label{kpnntab}
\end{table}
Note that the ranges given as the standard model
predictions in Table \ref{kpnntab} arise from our, at present,
limited knowledge of standard model parameters (CKM), and not
from intrinsic uncertainties in calculating 
the branching ratios.

\begin{figure}[t]
\hspace*{3cm}\epsfig{figure=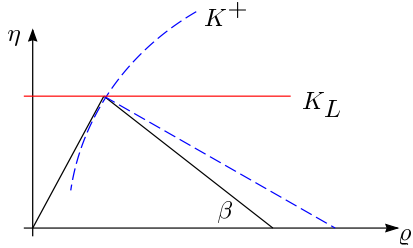,width=8.5cm,height=7.1cm}
\vspace*{-3.5cm}
\caption{Unitarity triangle from $K^+\to\pi^+\nu\bar\nu$ and
$K_L\to\pi^0\nu\bar\nu$.}
\label{fig:utkpn}
\end{figure}

With a measurement of $B(K^+\to\pi^+\nu\bar\nu)$ and 
$B(K_L\to\pi^0\nu\bar\nu)$ available very interesting phenomenological
studies could be performed. 
For instance, $B(K^+\to\pi^+\nu\bar\nu)$ and 
$B(K_L\to\pi^0\nu\bar\nu)$ together determine the unitarity triangle
(Wolfenstein parameters $\varrho$ and $\eta$) 
completely (Fig. \ref{fig:utkpn}). 
The expected accuracy with $\pm 10\%$ branching ratio measurements is
comparable to the one that can be achieved by CP violation studies
at $B$ factories before the LHC era \cite{BB6}.
The quantity $B(K_L\to\pi^0\nu\bar\nu)$ by itself offers probably the
best precision in determining $\mbox{Im} V^*_{ts}V_{td}$ or,
equivalently, the Jarlskog parameter
\begin{equation}\label{jcp}
J_{CP}=\mbox{Im}(V^*_{ts}V_{td}V_{us}V^*_{ud})=
\lambda\left(1-\frac{\lambda^2}{2}\right)\mbox{Im}\lambda_t
\end{equation}
The prospects here are even better than for $B$ physics at the LHC.
As an example, let us assume the following results will be
available from B physics experiments
\begin{equation}\label{lhcb}
\sin 2\alpha=0.40\pm 0.04\quad \sin 2\beta=0.70\pm 0.02\quad
V_{cb}=0.040\pm 0.002
\end{equation}
The small errors quoted for $\sin 2\alpha$ and $\sin 2\beta$ from
CP violation in $B$ decays require precision measurements at the LHC.
In the case of $\sin 2\alpha$ we have to assume in addition that the
theoretical problem of `penguin-contamination' can be resolved.
These results would then imply
$\mbox{Im}\lambda_t=(1.37\pm 0.14)\cdot 10^{-4}$.
On the other hand, a $\pm 10\%$ measurement 
$B(K_L\to\pi^0\nu\bar\nu)=(3.0\pm 0.3)\cdot 10^{-11}$ together with
$m_t(m_t)=(170\pm 3) GeV$ would give
$\mbox{Im}\lambda_t=(1.37\pm 0.07)\cdot 10^{-4}$. If we are optimistic
and take $B(K_L\to\pi^0\nu\bar\nu)=(3.0\pm 0.15)\cdot 10^{-11}$,
$m_t(m_t)=(170\pm 1) GeV$, we get
$\mbox{Im}\lambda_t=(1.37\pm 0.04)\cdot 10^{-4}$, a remarkable
accuracy. The prospects for precision tests of the standard model
flavour sector will be correspondingly good.

The charged mode $K^+\to\pi^+\nu\bar\nu$ is still being 
studied by Brookhaven experiment E787, which
will be followed by a successor experiment, E949 \cite{E949}. 
Recently, a new experiment, CKM \cite{CKM}, has been proposed to measure 
$K^+\to\pi^+\nu\bar\nu$ at the Fermilab Main Injector,
studying $K$ decays in flight.
Plans to investigate this process also exist at KEK for
the Japan Hadron Facility (JHF) \cite{JHFS}.

The neutral mode, $K_L\to\pi^0\nu\bar\nu$, is currently
pursued by KTeV. 
For $K_L\to\pi^0\nu\bar\nu$ a model
independent upper bound can be infered from the experimental result
on $K^+\to\pi^+\nu\bar\nu$, which at present is
stronger than the direct experimental limit \cite{GN}.
It is given by $B(K_L\to\pi^0\nu\bar\nu)< 4.4 B(K^+\to\pi^+\nu\bar\nu)
< 2\cdot 10^{-9}$. At least this sensitivity will have to be achieved
before new physics is constrained with $B(K_L\to\pi^0\nu\bar\nu)$.
Concerning the future of $K_L\to\pi^0\nu\bar\nu$ experiments, 
a proposal exists at Brookhaven (BNL E926) to measure this decay at 
the AGS with a sensitivity of ${\cal O}(10^{-12})$ \cite{E926}.
There are furthermore plans to pursue this mode with comparable
sensitivity at Fermilab \cite{KAMI} and KEK \cite{JHFI}.
Prospects for $K_L\to\pi^0\nu\bar\nu$ at a $\phi$-factory were
discussed in \cite{BCI}.

\subsection{$K_L\to\pi^0e^+e^-$}

The electric charge of the leptons and the resulting interaction
with photons make the decay $K_L\to\pi^0e^+e^-$ more complicated
than $K_L\to\pi^0\nu\bar\nu$.
The short-distance dominated part of the $K_L\to\pi^0e^+e^-$
amplitude can be analyzed using OPE and the renormalization group
in analogy to the case of the $\Delta S=1$ effective Hamiltonian.
This approach is reviewed in \cite{BBL}.
Here we would like to give a qualitative discussion, which
highlights the characteristic points and also summarizes the main
differences between $K_L\to\pi^0e^+e^-$ and $K_L\to\pi^0\nu\bar\nu$.

The basic diagrams for $K_L\to\pi^0e^+e^-$ are similar to those
for $K_L\to\pi^0\nu\bar\nu$, except that also a photon can be
exchanged instead of the $Z$ boson in the penguin diagrams
(see Fig. \ref{fig:kpnn}).
Taking the GIM mechanism into account, the structure of the effective
Hamiltonian reads
\begin{equation}\label{heffsim}
{\cal H}_{eff}\sim \lambda_t\, (F_t-F_c)+\lambda_u\, (F_u-F_c)
\end{equation}
Recalling that we can write $K_L\approx K_2+\varepsilon K_1$,
where $K_2$ ($K_1$) is the CP odd (CP even) neutral kaon state,
the decay amplitude takes the form

\begin{flushleft}
$A(K_L\to\pi^0 f\bar f)\sim $\\
$\ {\rm Im}\lambda_t (F_t-F_c) +
\varepsilon\left[{\rm Re}\lambda_t(F_t-F_c)+
{\rm Re}\lambda_u(F_u-F_c)\right]$\\
\ \ \ \ \\
{\small $\  10^{-4}\cdot \  1\ \ \ \ \ \ \  +
10^{-3}\left[ 10^{-4}\cdot 1 \ \ \ \ \ \ \ \ + 10^{-1}\cdot
\left\{\begin{array}{cl}10^{-3} & f=\nu\\ 1 & f=e
\end{array}\right. \right]$}
\end{flushleft}

\noindent
Here we have kept the expression general to allow for the cases
$f=e$ and $\nu$. We have assumed the structure of the
short-distance Hamiltonian for this exercise, although this is not
strictly correct for the parts that are sensitive to long-distance
physics. However it is sufficient for the qualitative argument
we would like to make. We recall a few points from the discussion
in the section on $K\to\pi\nu\bar\nu$.

First, as we have seen, the $K_2$ component yields an amplitude
proportional to the imaginary parts of the CKM elements.
Correspondingly, the $K_1$ amplitude (opposite CP),
which is multiplied by $\varepsilon$, gives the real parts.
Second, we need to consider that the GIM mechanism is hard for
$f=\nu$ and soft for $f=e$. This implies
$F_t\sim 1$, $F_c\sim 10^{-3}$, $F_u\sim 10^{-5}$ for $f=\nu$,
and $F_{u,c,t}\sim 1$ for $f=e$. 
Third, we determine the hierarchy of the CKM elements.
Putting this information together, we find the 
order-of-magnitude estimates shown above for the various terms.

For $f=\nu$ we recover what we already know from the previous
section: The amplitude is purely from direct CP violation
(the $\varepsilon$-component is negligible) and it is dominated 
by short-distance physics (only $F_t$, $F_c$ contribute).
For $f=e$ we can read off: Both direct and indirect CP violation
contribute at the same order ($\sim 10^{-4}$) and the latter 
component is determined by long-distance dynamics ($F_u$).

In addition to what we have discussed so far, a long distance
dominated, CP conserving amplitude with two-photon intermediate state
can contribute.  
\begin{figure}[t]
\hspace*{3cm}\epsfig{figure=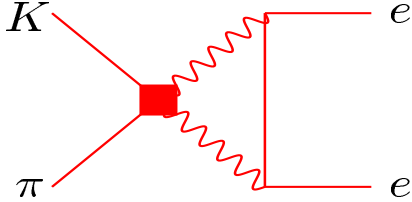,width=6cm,height=3cm}
\caption{CP conserving two-photon contribution to
$K_L\to\pi^0 e^+e^-$.}
\label{fig:kpeecpc}
\end{figure}
Although it is of higher order in the electromagnetic coupling,
it can potentially compete with the other contributions because
those are suppressed by CP violation.
Treating $K_L\to\pi^0e^+e^-$
theoretically one is thus faced with the need to disentangle three
different contributions of roughly the same order of magnitude.
\begin{itemize}
\item
Direct CP violation: This amplitude is short-distance in character,
theoretically clean and has been analyzed at next-to-leading order
in QCD \cite{BLMM}. Taken by itself this mechanism leads to a
$K_L\to\pi^0e^+e^-$ branching ratio of $(4.5\pm 2.6)\cdot 10^{-12}$
within the standard model.
\item
Indirect CP violation: This part is given by
$\sim\varepsilon\cdot A(K_S\to\pi^0e^+e^-)$. The $K_S$ amplitude is
dominated by long distance physics and has been investigated in
chiral perturbation theory \cite{EPR,BRP,DG,DEIP}. 
Due to unknown counterterms
that enter this analysis a reliable prediction is not possible at
present. The situation would improve with a measurement of
$B(K_S\to\pi^0e^+e^-)$, which could become possible at 
the CERN experiment NA48 or with KLOE at DA$\Phi$NE,
the Frascati $\Phi$-factory.
Present day estimates for $B(K_L\to\pi^0e^+e^-)$ due to indirect
CP violation alone allow values of
$10^{-12}$ -- $10^{-10}$.
\item
The CP conserving two-photon contribution is again long-distance
dominated. 
It has been analyzed by various authors \cite{DG,CEP,HS}.
The estimates are typically a few $10^{-12}$. Improvements in this sector
might be possible by further studying the related decay
$K_L\to\pi^0\gamma\gamma$ whose branching ratio has already been
measured to be $(1.7\pm 0.1)\cdot 10^{-6}$. 
\end{itemize}

Originally it had been hoped that the direct CP violating
contribution would be dominant. 
It is possible that the CP conserving part is not too important.
However, this is unlikely to be true for the amplitude from
indirect CP violation. Experimental input on $K_S\to\pi^0e^+e^-$
will be indispensable to solve this problem. We also mention
that the CP violating contributions interfere with a relative
phase, which is known up to a sign ambiguity. The CP conserving
part simply adds incoherently to the decay rate.

Besides the theoretical issues, $K_L\to\pi^0e^+e^-$ is also 
challenging
from an experimental point of view. The expected branching ratio 
is even smaller than for $K_L\to\pi^0\nu\bar\nu$. Furthermore a
serious irreducible physics background from the radiative mode
$K_L\to e^+e^-\gamma\gamma$ has been identified, which poses additional
difficulties \cite{LV}. A background subtraction seems necessary,
which should be possible with enough events. 
Additional information could in principle also be gained by studying
the electron energy asymmetry \cite{DG,HS} or the time evolution
\cite{DG,LIT,KP}.

\subsection{$K_L\to\mu^+\mu^-$}

\begin{figure}[t]
\hspace*{3cm}\epsfig{figure=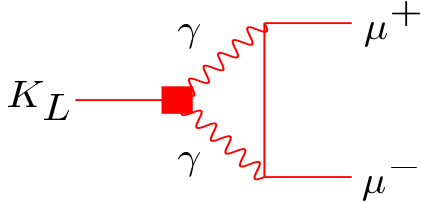,width=6cm,height=3cm}
\caption{The dominant contribution to $K_L\to\mu^+\mu^-$.}
\label{fig:klmm}
\end{figure}

$K_L\to\mu^+\mu^-$ receives a short distance contribution from
Z-penguin and W-box graphs similar to $K\to\pi\nu\bar\nu$. This
part of the amplitude is sensitive to the Wolfenstein parameter
$\varrho$. In addition $K_L\to\mu^+\mu^-$ proceeds through a
long distance contribution with the two-photon intermediate state,
which actually dominates the decay completely (Fig. \ref{fig:klmm}). 
The long distance
amplitude consists of a dispersive ($A_{dis}$) and an absorptive
contribution ($A_{abs}$). The branching fraction can thus be written
\begin{equation}\label{bklma}
B(K_L\to\mu^+\mu^-)=|A_{SD}+A_{dis}|^2 + |A_{abs}|^2
\end{equation}
Using $B(K_L\to\gamma\gamma)$ it is possible to extract \cite{LV}
$|A_{abs}|^2=(7.1\pm 0.2)\cdot 10^{-9}$.
$A_{dis}$ on the other hand cannot be calculated accurately at 
present \cite{GDP,KPPR,GV,DIP}.
This is rather unfortunate, in particular since 
$B(K_L\to\mu^+\mu^-)$ has already
been measured with very good precision
\begin{equation}\label{bklmex}
B(K_L\to\mu^+\mu^-)=
        (7.18\pm 0.17)\cdot 10^{-9} \ \mbox{BNL 871 \cite{AMB}}
\end{equation}
Interestingly, the absorptive contribution essentially
saturates the total rate. For comparison we note that \cite{AJB99} 
$B(K_L\to\mu^+\mu^-)_{SD}=|A_{SD}|^2=(0.9\pm 0.4)\cdot 10^{-9}$ is the
expected branching ratio in the standard model based on the
short-distance contribution alone. Because $A_{dis}$
is largely unknown, $K_L\to\mu^+\mu^-$ is at present not a very
useful constraint on CKM parameters.


\section{Flavour Physics with Charm}

\subsection{Rare $D$ Decays}

Weak decays of charmed particles, $D$ mesons in particular,
can also be used to probe the physics of flavour.
Due to the characteristic pattern of standard model
weak interactions, 
rare processes with $D$ mesons 
are markedly different from their kaon counterparts.
First of all, the charm quark mass $m_c\approx 1.4$ GeV is
considerably bigger than the strange quark mass, and also the QCD
scale $\Lambda_{QCD}$. 
Sometimes methods from the theory of heavy quarks can be employed
for charmed particles, although they are less reliable than in
the case of the much heavier $B$ mesons (see the lectures by
Falk in this volume for a general discussion in the context of
$B$ physics). This situation makes a theoretical treatment
of $D$ decays more difficult than for the heavy $B$ mesons on one
hand, and for the light kaons on the other.

More important, however, is the fact that charm is a quark with
weak isospin $T_3=+1/2$, in contrast to $s$ and $b$. For this reason
the virtual particles appearing in FCNC loop diagrams are the
down-type quarks $d$, $s$, $b$, rather than $u$, $c$, $t$ familiar
from $K$ and $B$ physics.
Examples of rare $D$ decays are
\begin{equation}\label{rared}
D\to\rho\gamma,\ D\to\pi l^+l^-,\ D\to\mu^+\mu^-,\
D\to\gamma\gamma,\ D\to\pi\nu\bar\nu
\end{equation}
They proceed through $c\to u$ FCNC transitions
and their $\Delta C=1$ amplitudes have the generic form
\begin{eqnarray}\label{fdsb}
&& V^*_{cd}V_{ud} F_d +V^*_{cs}V_{us} F_s +V^*_{cb}V_{ub} F_b 
\nonumber \\
&& = V^*_{cs}V_{us} (F_s - F_d) +V^*_{cb}V_{ub} (F_b - F_d)
\end{eqnarray}
where $F_i$ is the amplitude with internal quark $i=d$, $s$, $b$.
A potentially large contribution could come from $F_b$, which
can have a quadratic mass dependence $\sim m^2_b/M^2_W$. This exceeds
the contribution from light flavours in the loop, but it is still
much smaller than virtual top effects. In addition, the $b$-quark
contribution is very strongly CKM suppressed since 
$V^*_{cb}V_{ub}\sim\lambda^5$. On the other hand, 
$V^*_{cs}V_{us}\sim\lambda$ is much larger, but this term is
multiplied by $F_s-F_d$. The latter is non-zero only through the effects of
$SU(3)$ breaking and therefore receives a strong GIM suppression.  
Moreover, the light-quark loops $F_s$, $F_d$ are sensitive to 
nonperturbative QCD dynamics.
Additional long-distance mechanisms, different from $F_s$, $F_d$,
can also become important.

An example is \cite{SP} $D^0\to\mu^+\mu^-$. Here the amplitude
in (\ref{fdsb}) yields a tiny branching fraction of $\sim 10^{-19}$.
Alternatively the decay can proceed, for instance, via the long-distance
mechanism $D^0\to K^0\to\mu^+\mu^-$, with an off-shell kaon.
Estimates of this and similar
sources give together \cite{SP} $B(D^0\to\mu^+\mu^-)\sim 10^{-15}$.
This still leaves a window for the discovery of potential
new physics effects below the current experimental limit of \cite{PDG}
$B(D^0\to\mu^+\mu^-)_{exp} < 4\cdot 10^{-6}$.

\subsection{$D^0$--$\bar D^0$ Mixing}

A further interesting probe of the flavour sector is
$D^0$--$\bar D^0$ mixing, a process with $\Delta C=2$.
Recent measurements from the CLEO and FOCUS collaborations have
stimulated the interest in this observable. A detailed discussion
of these results and a list of references can be found in
\cite{BGLNP}. 

$D^0$--$\bar D^0$ mixing can be searched for in the time-dependent
analysis of hadronic two-body modes. We may distinguish three types of
decays,
\begin{eqnarray}\label{cfsd}
\mbox{Cabibbo favoured (CF):} && D^0\to K^-\pi^+ \\
\mbox{singly Cabibbo suppressed (SCS):} && D^0\to K^+ K^- \\
\mbox{doubly Cabibbo supressed (DCS):} && D^0\to K^+\pi^- 
\end{eqnarray}
In powers of the Wolfenstein parameter $\lambda$, the CF amplitude
$c\to su\bar d$ is of order $\lambda^0$, the SCS amplitude
$c\to su\bar s$ of order $\lambda$, and the DCS amplitude
$c\to du\bar s$ of order $\lambda^2$, which establishes a clear
hierarchy in the decay rates. The classification applies of course
also to the charge-conjugated modes
$\bar D^0\to K^+\pi^-$ (CF), $\bar D^0\to K^+ K^-$ (SCS) and
$\bar D^0\to K^-\pi^+$ (DCS).

The framework for describing $D^0$--$\bar D^0$ mixing is analogous
to the case of $K^0$--$\bar K^0$ mixing discussed in sec. 4.1.
The mass eigenstates are
\begin{equation}\label{d12ddb}
D_{1,2}=p D\pm q\bar D\qquad\qquad CP\cdot D=-\bar D
\end{equation}
We introduce the convenient definitions ($\Gamma$ is the 
average total decay rate)
\begin{equation}\label{xyxi}
x=\frac{M_2-M_1}{\Gamma}\quad y=\frac{\Gamma_2-\Gamma_1}{2\Gamma}
\quad \xi=\frac{p}{q}
\frac{\langle K^+\pi^-|D\rangle}{\langle K^+\pi^-|\bar D\rangle}
\end{equation}
$x$, $y$, $\xi$ are small quantities of order $\lambda^2$.
The precise calculation of $x$ and $y$ is difficult for the reasons
mentioned above (GIM suppression, long-distance sensitivity).
Analyses based on hadronic estimates ($K$, $\pi$ intermediate states)
or the heavy-quark expansion (quark-level calculation) lead to a
standard model expectation of
\begin{equation}\label{xynum}
x,\ y \stackrel{<}{_\sim} 10^{-4}
\end{equation}

We then consider the time dependent decay $\Gamma(D^0(t)\to K^+\pi^-)$.
The state $D^0(t)$ is obtained by solving the Schr\"{o}dinger
equation for the $D^0$--$\bar D^0$ two-state system with the initial
condition $D^0(t=0)=D^0$.
The solution reads
\begin{eqnarray}\label{dtkp}
&&\Gamma(D^0(t)\to K^+\pi^-)=e^{-\Gamma t}\left|\frac{q}{p}\right|^2
\Gamma(\bar D^0\to K^+\pi^-) \nonumber \\
&& \hspace*{1cm}\cdot\left[|\xi|^2+(y\,{\rm Re}\xi+x\,{\rm Im}\xi)\Gamma t
+\frac{1}{4}(x^2+y^2)(\Gamma t)^2\right]
\end{eqnarray}
To obtain this result we have used the following approximations.
We expand in the small quantities $\xi$, $x$, $y$, assumed to be
of the same order (of order $x$, say), and drop terms of
${\cal O}(x^4)$ and higher. We also require $\Gamma t$ to be at most
of ${\cal O}(1)$. This is the range that is relevant experimentally,
since the $D$ mesons will have decayed once $\Gamma t$ becomes
too large.

The form of (\ref{dtkp}) is not hard to interpret.
If there is no mixing, $x=y=0$, only the first term inside the
square brackets survives. It simply describes the exponentially
decaying rate of the DCS process $D^0\to K^+\pi^-$. Even if
$x$, $y\not= 0$, at time $t=0$ the DCS decay is the only effect.
On the other hand, if the small DCS decay was absent, $\xi=0$,
only the third term contributes. Then the $K^+\pi^-$
final state can arise only through mixing $D^0\to\bar D^0\to K^+\pi^-$.
The rate vanishes at $t=0$ because $D^0(t=0)=D^0$ and there was no
time yet for mixing.
The $(x\Gamma t)^2$ behaviour represents the onset of an oscillating
trigonometric function of which it is the remnant in our approximation.
Finally, the linear term $\sim\Gamma t$ is from the interference
of DCS decay and mixing.

Without the DCS mode, the observation of $K^+\pi^-$ from an initial
$D^0$ would be an unmistakable sign of mixing. The real situation
is more complicated. To demonstrate the presence of the mixing term,
the three contributions in (\ref{dtkp}) have to be disentangled, which
is possible in principle due to their different time dependence.
So far only the DCS component ($|\xi|^2$ term) has been unambiguously
identified. At present the signal for mixing is still compatible
with zero. It will be interesting to follow the future
experimental results on this issue. The detection of a substantial
value of $x$ would indeed be exciting evidence for
new physics.


\section{Summary and Outlook}

Kaon decays have played a key role in the development of the
standard model. Today kaon physics is a mature field.
A whole array of modern field theoretical techniques is at
our disposal to help us extract the underlying mechanisms.
Among these tools are perturbative quantum field theory,
including the perturbative treatment of QCD at short distances,
the operator product expansion, the renormalization group
and chiral perturbation theory.
While an impressive number of crucial insights has been already obtained
in the past, excellent opportunities continue to exist for
present and future studies:

\begin{itemize}
\item
Chiral perturbation theory constitutes a complementary handle
on the elusive nonperturbative dynamics of QCD at long distances.
This can be helpful to control long-distance contributions that
contaminate the short-distance physics that is of primary interest.
However, chiral perturbation theory, as a model-independent
framework for low-energy QCD, is also of considerable interest
in its own right. Typical processes that are studied in this context
are $K^+\to\pi^+ l^+l^-$, $K_L\to\pi^0\gamma\gamma$, 
or $K_S\to\gamma\gamma$. 
\item
CP violation in $K\to\pi\pi$ is still an important area of
current interest. Indirect CP violation measured by $\varepsilon$
is well determined experimentally and provides us with a valuable
CKM constraint. Direct CP violation, now established, but still under
further experimental investigation, gives an important qualitative
test of the standard model.
\item
Standard model precision tests will be possible with the
``golden'' decay modes $K^+\to\pi^+\nu\bar\nu$ and
$K_L\to\pi^0\nu\bar\nu$.
\item
Other opportunities of interest include decays as
$K_L\to\pi^0 e^+e^-$, $\mu$-polarization in $K^+\to\pi^0\mu^+\nu$,
among many other rare processes.
\item
Very direct and clean  probes for new physics are decays
that are forbidden in the standard model. Lepton flavour violating
modes as $K_L\to e\mu$, $K\to\pi\mu e$ are important examples.
\end{itemize}

In parallel to kaon physics many other observables, as provided
from decays of hadrons with beauty and charm, will be necessary
to get a reliable and complete picture of the physics of flavour
and its possible origins.

In these lectures, we have discussed selected examples
from the phe\-no\-me\-no\-lo\-gy of mesons with strangeness and charm.
We have also emphasized the theoretical framework
for an analysis of these processes, which is crucial to interpret
the experimental data and to extract the underlying physics. 
The coming years promise to be very fruitful for the study of
flavour physics and important discoveries are possible in the near future.

\section*{Acknowledgments}

I thank the organizers of TASI 2000 for inviting me to this
very interesting and pleasant Summer School and for their 
hospitality at Boulder. Thanks are also due to the students
for their active participation. I am grateful to Gino Isidori
and Jon Rosner for comments on the manuscript.

\end{document}